\documentclass[12pt]{article}
\usepackage{amsbsy}
\usepackage{amsmath}
\usepackage{amssymb}
\usepackage{amsthm}
\usepackage{float}
\usepackage{bm}
\usepackage{mathrsfs}
\usepackage{enumerate,enumitem}
\usepackage{multirow}
\usepackage{multicol}
\usepackage{array}
\usepackage{color}
\usepackage{epsfig}
\usepackage{graphicx}	
\usepackage{subfigure}
\usepackage{booktabs}
\usepackage{times}
\usepackage{algorithm}
\usepackage{algorithmic}
\usepackage{threeparttable}
\usepackage[title]{appendix}
\usepackage[T1]{fontenc}
\usepackage[utf8]{inputenc}
\usepackage{natbib}
\usepackage{hyperref}
\usepackage{chngcntr}
\usepackage{caption,subcaption}

\newtheorem{theorem}{Theorem}

\newtheorem{remark}{Remark}

\newtheorem{assumption}{Assumption}

\newcommand{\blind}{1}

\date{}
\begin{document}

\def\spacingset#1{\renewcommand{\baselinestretch}%
{#1}\small\normalsize} \spacingset{1}


\if1\blind
{
  \title{\bf Subgroup learning in functional regression models under the RKHS framework}
  \author{Xin Guan\\
    School of Statistics and Mathematics, Zhongnan University of Economics and Law\\
    and \\
    Yiyuan Li\\
    School of Statistics and Management, Shanghai University of Finance and Economics\\
    and \\
     Xu Liu\thanks{Liu's work was supported partially by the National Natural Science Foundation of China (12271329,72331005) and the Program for Innovative Research Team of SUFE, the Shanghai Research Center for Data Science and Decision Technology, and the Open Research Fund of Yunnan Key Laboratory of Statistical Modeling and Data Analysis, Yunnan University.}\hspace{.2cm}\\
     School of Statistics and Management, Shanghai University of Finance and Economics\\
     and \\
     Jinhong You\thanks{The research of You was supported in part by the National Natural Science Foundation of China (11971291).}\hspace{.2cm} \\
     School of Statistics and Management, Shanghai University of Finance and Economics 
     }
  \maketitle
} \fi

\if0\blind
{
  \bigskip
  \bigskip
  \bigskip
  \begin{center}
    {\LARGE\bf Subgroup learning in functional regression models under the RKHS framework}
\end{center}
  \medskip
} \fi

\bigskip
\begin{abstract}
Motivated by the inherent heterogeneity observed in many functional or imaging datasets,
this paper focuses on subgroup learning in functional or image responses.  
While change-plane analysis has demonstrated empirical success in practice, the existing methodology is confined to scalar or longitudinal data.
In this paper, we propose a novel framework for estimation, identifying, and testing the existence of subgroups in the functional or image response through the change-plane method.
The asymptotic theories of the functional parameters are established based on the vector-valued Reproducing Kernel Hilbert Space (RKHS), and the asymptotic properties of the change-plane estimators are derived by a smoothing method since the objective function is nonconvex concerning the change-plane.
A novel test statistic is proposed for testing the existence of subgroups, and its asymptotic properties are established under both the null hypothesis and local alternative hypotheses.
Numerical studies have been conducted to elucidate the finite-sample performance of the proposed estimation and testing algorithms.
Furthermore, an empirical application to the COVID-19 dataset is presented for comprehensive illustration.
\end{abstract}

\noindent%
{\it Keywords:}   Change-plane analysis; Functional response; Heterogeneity; Hypothesis testing; Reproducing kernel Hilbert space.
\vfill

\newpage
\spacingset{1.5} 

\section{Introduction}\label{sec1}
In recent decades, functional data analysis has found widespread application in diverse fields, including finance \citep{fan2015}, medical science \citep{huang2020}, and neuroimaging \citep{zhu2023}. 
A growing body of literature has emerged in this area, with much of the research focused on functional regression models that consider the average effects of regressors. For comprehensive reviews, refer to \cite{ramsay2005,wang2016}.
However, the traditional functional regression model may yield biased estimates and inferences if population heterogeneity is neglected.

\begin{figure}[!htbp]
	\centering
	\subfigure[]{\includegraphics[width=0.4\linewidth]{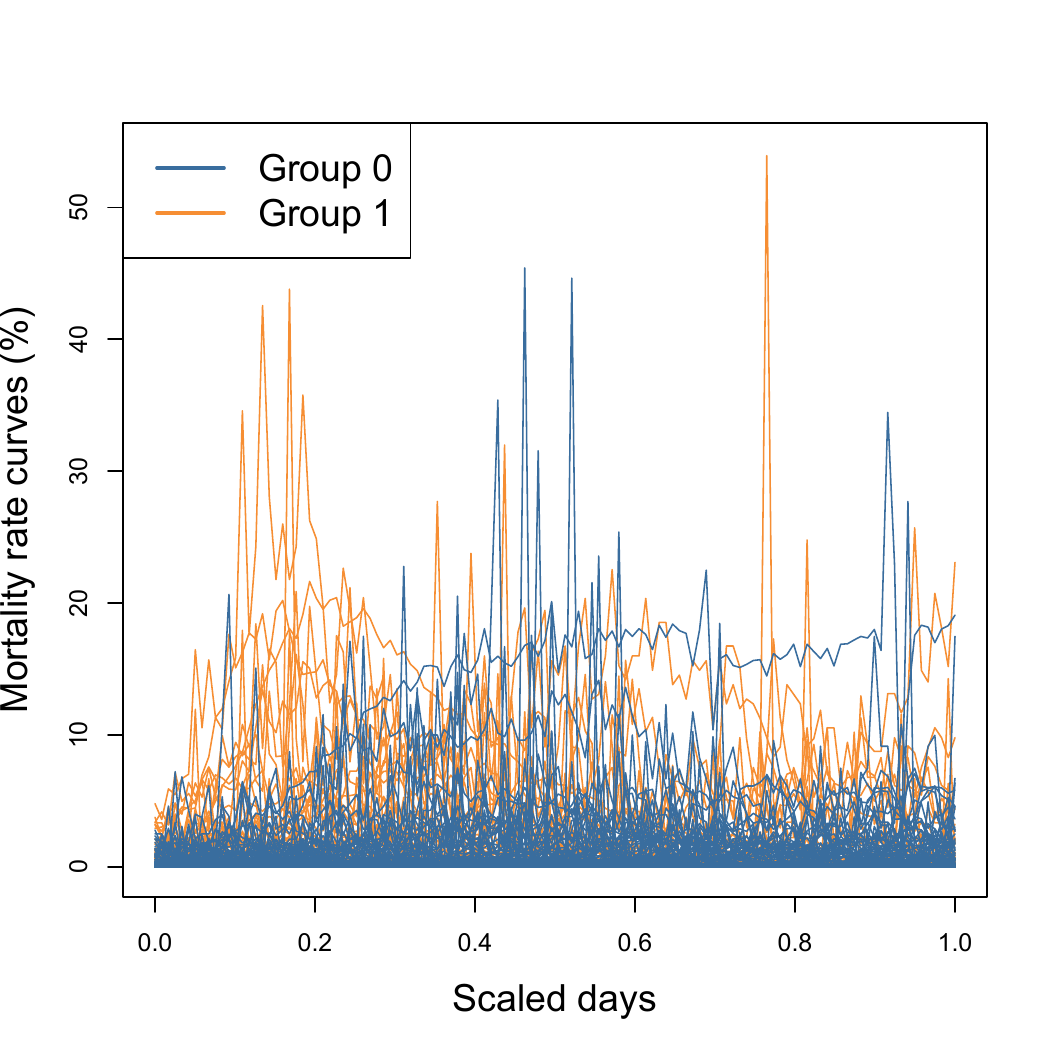}}\hspace{12pt}
	\subfigure[]
	{\includegraphics[width=0.4\linewidth]{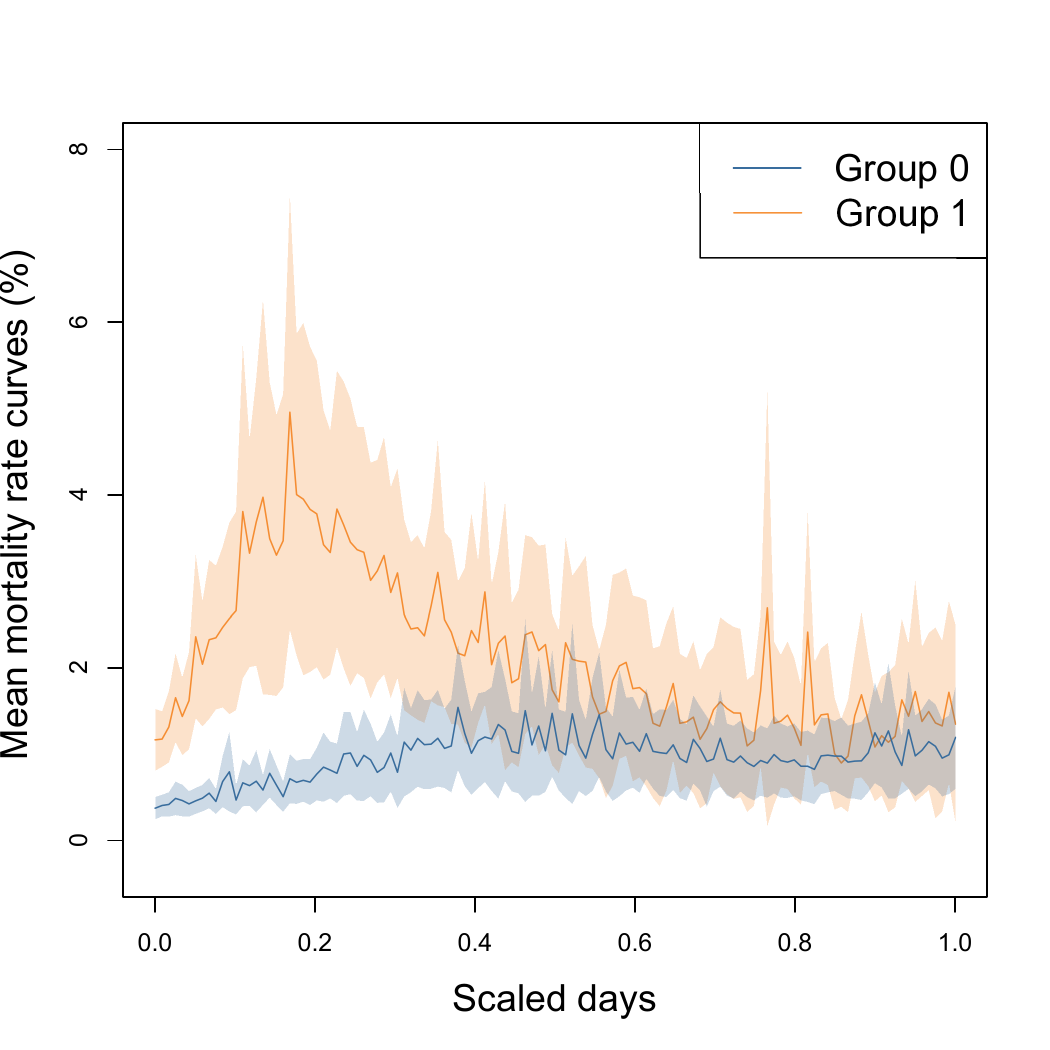}}
	\captionsetup{width=\textwidth}
	\caption{(a) The mortality rate curves ($\times 100\%$) of COVID-19 data for 137 countries with two subgroups, where ``Group 0" and ``Group 1" determine two subgroups, see details in Section \ref{sec6}. (b) The mean mortality rate curves ($\times 100\%$) and their 95\% pointwise confidence bands for two subgroups in the COVID-19 dataset.}
	\label{fig1}
\end{figure}

The motivation for this paper comes from analyzing mortality rates in the COVID-19 dataset, sourced from the World Health Organization (\url{http://covid19.who.int/}). This dataset, which is a typical example of functional data, contains mortality rate curves from 137 countries over 120 days after each country reached 100 confirmed cases, along with information on population aging and medical care conditions.
To explore the heterogeneity in mortality rate curves, we conduct a subgroup analysis, dividing the 137 countries into two subgroups (as defined later in Section \ref{sec6}).
Figure \ref{fig1}(a) displays the mortality rate curves of 137 countries, while Figure \ref{fig1}(b) presents the mean mortality rate curves along with the 95\% pointwise confidence bands for each subgroup. The pronounced difference in mean mortality rate curves between these subgroups, illustrated in Figure \ref{fig1}(b), is supported by a p-value of 0.002 derived from our proposed testing method, providing strong evidence for the existence of subgroups. Furthermore, a two-sample test for functional data, developed by \cite{qiu2021}, produces a p-value of less than 0.01, further substantiating the significant differences in mean mortality rate curves between the two subgroups.
Therefore, ignoring this heterogeneity could lead to model misspecification.

The infinite-dimensional nature of functional data poses significant challenges for subgroup learning. Despite its importance, subgroup learning for functional responses has not been extensively studied in the literature, with most existing research focused primarily on subgroup identification methods.
For instance, \cite{wang2016Mixture} and \cite{jiang2022} explored clustering analysis with functional responses using a mixed-Gaussian model, allowing regression structures to vary across latent subgroups. However, the mixture model imposes strict distributional assumptions and is computationally intensive. Additionally, \cite{collazos2023} developed a K-means algorithm to identify latent groups of functional responses, but this approach lacks a solid statistical foundation.
In contrast, change-plane analysis has proven to be a powerful alternative for subgroup identification and testing, with successful applications in fields such as precision medicine \citep{songrui2017} and economic structures with systemic breaks \citep{hansen2000, zhangyy2021}.
However, available change-plane methods have mainly concentrated on scalar responses, including continuous \citep{seolinton2005}, binary \citep{huang2020}, and longitudinal \citep{wei2023}, leaving functional data largely underexplored. 
Our aim is to tackle these difficulties and fill the gap in the study of subgroup learning for functional data.

Herein, we focus on the change-plane analysis in the functional response regression model to learn the subgroups in functional responses with a set of scalar predictors. Let $\mathcal{S}$ be a bounded domain and $Y(s)$ be the functional response process for $s \in \mathcal{S}$, the change-plane model is
\begin{align}\label{1}
	Y(s)={\bm X}^{\mathrm{T}} {\bm \beta}(s)+\tilde{\bm X}^{\mathrm{T}} {\bm \delta}(s) I\left({\bm Z}^{\mathrm{T}} {\bm \psi} > 0\right)+\nu(s),
\end{align}
where $I(\cdot)$ is an indicator function, ${\bm X} \in \mathbb{R}^{p}$ denotes scalar predictors, $\tilde{\bm X} \in \mathbb{R}^{d}$($1 \leq d \leq p$) is a subset of ${\bm X}$, ${\bm Z} \in \mathbb{R}^{q+1}$ is the change-plane variable, ${\bm \psi} \in \mathbb{R}^{q+1}$ is the unknown grouping parameter, $\bm \beta(s)=(\beta_1(s), \cdots, \beta_p(s))^{\mathrm{T}}$ and $\bm \delta(s)=(\delta_1(s), \cdots, \delta_d(s))^{\mathrm{T}}$ are unknown functions, $\nu(s)$ characterizes individual variations and is assumed to be a stochastic process with mean zero and covariance function $\Lambda(s,t) = \text{cov}(\nu(s),\nu(t))$.
The model (\ref{1}) establishes a hyperplane through a linear combination of covariates ${\bm Z}$, and identifies distinct subgroups by whether ${\bm Z}^{\mathrm{T}} {\bm \psi}$ exceeds zero.
Specifically, the function ${\bm \beta}(s)$ measures the  main impact of ${\bm X}$ on the functional response process, whereas the function $\bm \delta(s)$ describes the change-plane effect, representing the heterogeneous influence of $\tilde{\bm X}$ in subgroups. When $\bm \delta(s) \equiv 0$, there is no heterogeneous in the population and the proposed model simplifies to the classical function-on-scalar regression, see \cite{zhu2012}, \cite{zhu2014} and \cite{lijialiang2017}.

Fitting model (\ref{1}) requires estimating both functional parameters ${\bm \beta}(s), {\bm \delta}(s)$ and the real-valued parameter $\bm \psi$. The popular method for approximate functional parameters is to use truncated expansions of basis functions, such as functional principal component analysis (FPCA), B-splines, or Fourier basis functions \citep{yao2005,hall2007,liu2021}. 
Although these methods are effective in dimensionality reduction, truncation can result in imprecise estimates, making subsequent statistical testing more difficult \citep{wahba1990}.
Following the idea in \cite{caiyuan2011} and \cite{shang2015}, we employ a roughness regularization method in the reproducing kernel Hilbert space (RKHS) framework to avoid these limitations.
The objective function for estimating $\bm \psi$ is nonconvex. Previous research has demonstrated that a direct estimator of $\bm \psi$ yields a nonstandard limiting distribution, making statistical inference overly complex \citep{yu2020, zhangyy2021}. To address this challenge, we introduce a smoothed estimator by applying a smoothing technique to the indicator function within the change-plane structure.
Since the solution lacks a closed-form expression, we propose an iterative algorithm to obtain the estimates. Additionally, we improve the estimation efficiency by accounting for spatial dependency in the functional responses.

In subgroup identification, false-positive grouping results may arise without sound hypothesis testing procedures.
For statistical inference in model (\ref{1}), one challenge is that $\bm \delta(\cdot)$ is not identifiable under the null hypothesis, and another is that the test statistic must be adaptable across the entire domain.
Previous work has investigated change-plane inference for scalar responses with non-dynamic subgroup effects.
For example, \cite{songrui2017} proposed a doubly robust score test statistic for subgroup testing with a continuous scalar response and a non-dynamic enhanced treatment effect. 
\cite{huang2020} developed a subgroup testing procedure based on maximum likelihood ratio statistics with a binary response.
More literature can refer to \cite{kang2017} and  \cite{kang2022}. 
However, due to the flexibility of model (\ref{1}), which allows subgroup effects to vary over time or spatial points, developing an appropriate method for subgroup testing becomes challenging. In this paper, we introduce a novel supremum of squared score test statistic, specifically adapted for dynamic subgroup effects, to determine the existence of subgroups in model (\ref{1}).
We initially derive a score estimation equation by computing the Fr\'{e}chet derivative of ${\bm \delta}(s)$ in $L_2$ space, 
followed by integrating over the domain of the observed functional data. To approximate the critical value, we develop a resampling procedure. Numerical studies demonstrate that the test statistics perform well in terms of both size and power.

To the best of our knowledge, this study is the first to explore the identification and testing of subgroups in functional data using change-plane analysis.
Our main contributions are summarized as follows. First,
the proposed change-plane model (\ref{1}) can efficiently estimate and identify subgroups within the functional responses.
Second, we develop a novel supremum of squared score test statistic is  to test the existence of subgroups in functional responses, and propose a resampling procedure to approximate the critical value of this test statistic. 
Third, we establish the asymptotic properties of the estimated functional parameters within the vector-valued RKHS framework. We also develop the asymptotic theory for the estimator of the grouping parameter, which attains a convergence rate of $h^{1/2}n^{-1/2}$ with $h$ is the bandwidth and $n$ is the sample size.
Moreover, we derive the asymptotic distributions of the proposed test statistics under both the null and local alternative hypotheses.

The remainder of the paper is structured as follows. In Section \ref{sec2}, we propose a regularized estimation approach for functional parameters in the Sobolev space and a smoothing method to estimate the grouping parameter. In Section \ref{sec3}, we develop the supremum of squared score test statistics for identifying the existence of subgroups and provide a resampling procedure to approximate the critical value. Section \ref{sec4} focuses on establishing the asymptotic properties of the functional estimators and the grouping estimator, as well as deriving the limiting distribution for the test statistics.
Section \ref{sec5} presents simulation studies to assess the performance of the proposed method. In Section \ref{sec6}, we provide an illustration of application to the COVID-19 dataset. Section \ref{sec7} concludes the paper, and all technical proofs are presented in the supplementary materials.

\section{Estimation Procedure}\label{sec2}
In this section, we present the estimation methodology for the change-plane model, which enables the identification of subgroups in functional responses.
In empirical applications, the process $\{Y(s), s \in \mathcal{S}\}$ is typically measured at specific locations with random errors. Let $\{(Y_{i}(s_m), {\bm X}_{i}, \tilde{\bm X}_i, {\bm Z}_{i}), i=1,\cdots,n, m=1,\cdots,M\}$ be the observations, the sample version of model (\ref{1}) can be written as
\begin{align}\label{2}
	Y_i(s_m)={\bm X}_i^{\mathrm{T}} {\bm \beta}(s_m)+\tilde{\bm X}_i^{\mathrm{T}} {\bm \delta}(s_m) I\left({\bm Z}_{i}^\mathrm{T} {\bm \psi} > 0\right)+\nu_i(s_m)+e_i(s_m),
\end{align}
where $Y_i(s_m)$ is the functional response of subject $i$ at location $s_m$, $\nu_i(s_m)$ is a realization of
process $\nu(s)$, and $e_i(s_m)$ is the additional measurement error with mean zero and covariance function $\mathcal{E}(s,t) = \text{cov}(e_i(s),e_i(t))I(s = t)$. Moreover, $\nu_i(s)$ and $e_i(s)$ are assumed to be mutually independent. Since the grouping parameter ${\bm \psi}$ in the indicator function is not identifiable, we impose an identifiability condition in model (\ref{2}). That is, we normalize the first element of ${\bm \psi}$ to one, while the remaining elements are denoted as ${\bm \gamma} \in \mathbb{R}^q$. Let $Z_{1}$ denote the first element of ${\bm Z}$, and ${\bm Z}_{2}$ represent the remaining elements of ${\bm Z}$, with the first element of ${\bm Z}_{2}$ acting as the intercept. Then ${\bm Z}^{\mathrm{T}} {\bm \psi} = Z_{1} + {\bm Z}_{2}^\mathrm{T} {\bm \gamma}$. This technique is similarly employed in \cite{seolinton2005} and \cite{zhangyy2021}. 
Next, we outline the estimation procedures for the unknown parameters.

\subsection{The RKHS Estimation}\label{sec2.1}
Denote ${\bm \theta}=({\bm \beta}(\cdot)^\mathrm{T},  {\bm \delta}(\cdot)^\mathrm{T})^\mathrm{T}$ and 
${\bm \eta} = ({\bm \theta}^\mathrm{T}, {\bm \gamma}^\mathrm{T})^\mathrm{T}$. Suppose that each component function of ${\bm \theta}$ belongs to the $\alpha$th order Sobolev space $\mathcal{H}^{(\alpha)}(\mathcal{S})$, which is abbreviate as $\mathcal{H}$ for simplicity
\begin{align*}
	\mathcal{H}^{(\alpha)}(\mathcal{S}) = &\left\{f : \mathcal{S} \rightarrow \mathbb{R} | f^{(j)} \text{ is absolutely continuous for } j=1, \cdots \alpha-1, f^{(\alpha)} \in L_2(\mathcal{S})\right\},
\end{align*}
where $f^{(j)}$ is the $j$th derivative of $f(\cdot)$, and $L_2(\mathcal{S})$ is the $L_2$ space defined in $\mathcal{S}$. As noted in \cite{caiyuan2011} \cite{cheng2015}, we assume that $\alpha > 1/2$, such that $\mathcal{H}$ is an RKHS. For simplicity, assume that the null space of $\mathcal{H}$ is $\{0\}$ \citep{wangxiao2020}.
Let $\mathcal{H}^{p+d}$ represent the full parameter space for ${\bm \theta}$.
Let $K(\cdot,\cdot): \mathcal{S} \times \mathcal{S} \rightarrow \mathcal{H}$ be the reproducing kernel function of $\mathcal{H}$. Common choices for $K(\cdot,\cdot)$ in practice include the polynomial kernel and the Gaussian kernel. For more details, see \cite{wahba1990} and \cite{li2007}.

The objective function of model (\ref{2}) is 
\begin{align*}
	\mathcal{L}_{nM}({\bm \eta})=\frac{1}{2nM} \sum_{i=1}^{n} \sum_{m=1}^{M}\left[Y_{i}\left(s_m\right)-{\bm X}_{i}^{\mathrm{T}} {\bm \beta}(s_m)-\tilde{\bm X}_i^\mathrm{T} {\bm \delta}(s_{m}) I(Z_{1i} + {\bm Z}_{2i}^\mathrm{T} {\bm \gamma} > 0)\right]^{2}.
\end{align*}
To achieve regularized estimation of the functional parameters, we consider a penalized loss function
\begin{align*}
	\mathcal{L}_{nM, \lambda}({\bm \eta})=\mathcal{L}_{nM}({\bm \eta})+ \frac{\lambda}{2}J(\bm \theta, \bm \theta),
\end{align*}
where $J(\bm \theta, \bm \theta)$ is a roughness penalty on $\bm \theta$, 
and $\lambda$ is a regularization parameter. Throughout this paper, we assume that $J(\bm \theta, \bm \theta) = \left\langle {\bm \theta}, {\bm \theta} \right\rangle_{K} =  \sum_{k=1}^p \|\beta_k\|_{K}^2 + \sum_{l=1}^d \|\delta_l\|_{K}^2$, where $\| \cdot \|_{K}$ is a seminorm in $\mathcal{H}$.

As \cite{yu2020} noted, a direct estimate of $\bm \gamma$ leads to a nonstandard limiting distribution. Therefore, we adopt a smoothing function $G(\cdot)$, such as the cumulative distribution function of a normal distribution, to approximate the indicator function.
The smoothed estimator of $\bm \eta$ is 
\begin{align}\label{5}
	\hat{\bm \eta} = \mathop{\arg\min}\limits_{\bm \eta} \mathcal{L}_{nM,\lambda}({\bm \eta};h),
\end{align}
where $\mathcal{L}_{nM,\lambda}({\bm \eta}; h) = \mathcal{L}_{nM}({\bm \eta}; h)  + {\lambda}J(\bm \theta, \bm \theta)/{2}$, with
\begin{align*}
	\mathcal{L}_{nM}({\bm \eta}; h)=\frac{1}{2nM} \sum_{i=1}^{n} \sum_{m=1}^{M}\left[Y_{i}\left(s_m\right)-{\bm X}_{i}^{\mathrm{T}} {\bm \beta}(s_m)-\tilde{\bm X}_i^\mathrm{T} {\bm \delta}(s_{m}) G_h\left(Z_{1i} + {\bm Z}_{2i}^\mathrm{T} {\bm \gamma}\right)\right]^{2},
\end{align*}
and $G_h(\cdot)=G(\cdot/h)$, $h \rightarrow 0$ is a bandwidth parameter.
There is no closed-form solution for minimizing the objective function (\ref{5}), therefore, we propose a profiled estimation method.
For a given ${\bm \gamma}$, suppose that $(\tilde{\bm \beta}_{\bm \gamma}, \tilde{\bm \delta}_{\bm \gamma})$ minimizes the smoothed objective function
\begin{align}\label{eq:estimate}
	(\tilde{\bm \beta}_{\bm \gamma}, \tilde{\bm \delta}_{\bm \gamma})= \mathop{\arg\min}\limits_{{\bm \beta}, {\bm \delta}} \mathcal{L}_{nM, \lambda}({\bm \eta}; h).
\end{align}
Then we can estimate the grouping parameter ${\bm \gamma}$ by 
\begin{align}\label{eq:gamma}
	\hat{\bm \gamma} = \mathop{\arg\min}\limits_{\bm \gamma}\mathcal{L}_{nM}(\tilde{\bm \beta}_{\bm \gamma}, \tilde{\bm \delta}_{\bm \gamma}, {\bm \gamma}; h).
\end{align}
The profiled estimate of the functional parameters is thus defined as $\hat{\bm \beta}(s) = \tilde{\bm \beta}_{\hat{\bm \gamma}}(s)$, $\hat{\bm \delta}(s) =\tilde{\bm \delta}_{\hat{\bm \gamma}}(s)$.

For practical implementation, by the representer theorem of \cite{wahba1990}, the profiled estimators in (\ref{eq:estimate}) can be written as $\tilde{\bm \beta}_{\bm \gamma}(s)=(\tilde{\beta}_{k,{\bm \gamma}}(s), 1 \leq k \leq p)^\mathrm{T}$ and $\tilde{\bm \delta}_{\bm \gamma}(s) = (\tilde{\delta}_{l,{\bm \gamma}}(s), 1 \leq l \leq d)^\mathrm{T}$ with
\begin{align}\label{7}
	\tilde{\beta}_{k, \bm \gamma}(s)= {\bm b}_{k}^\mathrm{T} {\bm K}_s, \quad  \tilde{\delta}_{l, \bm \gamma}(s)= {\bm c}_{l}^\mathrm{T} {\bm K}_s,
\end{align}
where ${\bm K}_{s} = (K(s,s_1),\cdots,K(s,s_M))^\mathrm{T}$, ${\bm b}_k=(b_{km}, 1 \leq m \leq M)^\mathrm{T}$ and ${\bm c}_l=(c_{lm},  1 \leq m \leq M)^\mathrm{T}$ are coefficients.
Denote ${\bm K}=({\bm K}_{s_1}, \cdots,{\bm K}_{s_M})$, ${\bm d}=({\bm b}^\mathrm{T}, {\bm c}^\mathrm{T})^\mathrm{T}$ with ${\bm b}=({\bm b}_1^\mathrm{T}, \cdots, {\bm b}_p^\mathrm{T})^\mathrm{T}$, and ${\bm c}=({\bm c}_1^\mathrm{T}, \cdots, {\bm c}_d^\mathrm{T})^\mathrm{T}$.
Given the format of the solution (\ref{7}), we can rewrite the (\ref{eq:estimate}) in finite form as
\begin{align*}
	\mathcal{L}_{nM,\lambda}({\bm d}, {\bm \gamma}; h)=&\frac{1}{2n M} \sum_{i=1}^{n} \sum_{m=1}^{M}\left[Y_{i}\left(s_{m}\right)-\sum_{k=1}^{p} X_{ik} {\bm K}_{s_m}^\mathrm{T} {\bm b}_{k}-\sum_{l=1}^{d} \widetilde{X}_{il} {\bm K}_{s_m}^\mathrm{T} {\bm c}_{l}G_h(Z_{1i} + {\bm Z}_{2i}^\mathrm{T} {\bm \gamma}) \right]^{2}\\
	&\quad \quad \quad +\frac{\lambda}{2}\sum_{k=1}^p {\bm b}_k^{\mathrm{T}} {\bm K} {\bm b}_k +\frac{\lambda}{2}\sum_{l=1}^d {\bm c}_l^{\mathrm{T}} {\bm K} {\bm c}_l. 
\end{align*}
Therefore, the RKHS method transforms the infinite-dimensional optimization problem into one in finite dimensions. 
Denote $\mathbb{Y}_i=(Y_i(s_1), \cdots, Y_i(s_M))^\mathrm{T}$, and $\Omega=\left(\begin{array}{ll}{\bm I}_p & \\ & {\bm I}_d \end{array}\right) \otimes {\bm K}$, where $\otimes$ is the Kronecker product and ${\bm I}_p$ is the $p$ dimensional identity matrix.
The iterative estimation procedure is outlined in Algorithm \ref{alg1}, and the tuning parameter is selected using the cross-validation method.

\begin{algorithm}[H] 
	\caption{Calculate RKHS estimators in Section \ref{sec2.1}}
	\label{alg1}
	\begin{algorithmic}[1]
		\STATE For an initial ${\bm \gamma}$, denote ${\bm N}_{i,\bm \gamma}= ({\bm X}_i^\mathrm{T},\tilde{\bm X}_i^\mathrm{T} G_h({Z_{1i} + {\bm Z}_{2i}^\mathrm{T} {\bm \gamma}})) \otimes {\bm K}$.
		Then $$\hat{\bm d}=\mathop{\arg\min}\limits_{\bm d}\mathcal{L}_{nM,\lambda}({\bm d}, {\bm \gamma}; h) = \left(\sum_{i=1}^n {\bm N}_{i,\bm \gamma}^\mathrm{T} {\bm N}_{i,\bm \gamma} + \lambda nM \Omega \right)^{-1}\left(\sum_{i=1}^n {\bm N}_{i,\bm \gamma}^\mathrm{T}\mathbb{Y}_i\right).$$
		\STATE Given the profiled estimators $\tilde{\beta}_{k,{\bm \gamma}}(s) = {\bm K}_{s}^\mathrm{T} \hat{\bm b}_k$ for $1 \leq k \leq p$, $\tilde{\delta}_{l,{\bm \gamma}}(s) = {\bm K}_{s}^\mathrm{T} \hat{\bm c}_l$ for $1 \leq l \leq d$ obtained from  step 1,  update ${\bm \gamma}$ by (\ref{eq:gamma}).
		\STATE Repeat steps 1-2 and iterate until convergence.%
	\end{algorithmic}
\end{algorithm}

\subsection{The Weighted Estimation}\label{sec2.4}
Spatial dependence plays a crucial role in functional data analysis \citep{zhu2014,lijialiang2017}. In this subsection, we propose a weighted estimation method that accounts for the spatial dependence of the functional response, thereby enhancing the initial estimates.

Let ${\bm \Phi} = \{\Phi(s_k,s_l)\}_{k,l=1,\cdots,M}$ be the covariance matrix for the functional responses observed at $M$ grid points, where $\Phi(s,t) = \Lambda(s,t)+\mathcal{E}(s,s)$. Let $y_i^{*}(s) = Y_i(s) -{\bm X}_i^\mathrm{T} \hat{\bm \beta}(s)-\tilde{\bm X}_i^\mathrm{T}\hat{\bm \delta}(s)G_h({Z}_{1i}+{\bm Z}_{2i}^{\mathrm{T}} \hat{\bm \gamma})$, then $ y_i^{*}(s) \approx \nu_i(s)+e_i(s)$.
Assume that $\nu_i \in \mathcal{H}$, then the estimator $\hat{\nu}_i$ can obtained by minimizing the following objective function
\begin{align*}
	\hat{\nu}_i = \mathop{\arg\min}\limits_{{\nu} \in \mathcal{H}} \left\{\frac{1}{M}\sum_{m=1}^M  [y_i^{*}(s_m) -\nu(s_m)]^2 + {\lambda} \|\nu\|^2_{K}\right\}.
\end{align*}
By the representer theorem, the solution has a finite form $\hat{\nu}_i(s)={\bm K}_{s}^\mathrm{T} \hat{\bm f}_i$ with $\hat{\bm f}_i = (\hat{f}_{i1},\cdots,\hat{f}_{iM})^\mathrm{T}$. Using the least squares method, we can obtain $\hat{\bm f}_i$ by
\begin{align*}
	\hat{\bm f}_i 
	&= \mathop{\arg\min}\limits_{{\bm f}_i} \left\{[\mathbb{Y}_i^{*} - {\bm K}{\bm f}_i]^\mathrm{T}[\mathbb{Y}_i^{*} - {\bm K}{\bm f}_i] + \lambda M {\bm f}_i^\mathrm{T} {\bm K} {\bm f}_i \right\}\\
	& =  ({\bm K} + \lambda M {\bm I})^{-1}\mathbb{Y}_i^{*},
\end{align*}
with $\mathbb{Y}_i^{*}=(y_i^{*}(s_1), \cdots, y_i^{*}(s_M))^\mathrm{T}$. Then the covariance matrix $\Lambda(s,t)$ can be estimated by the empirical covariance
\begin{align}\label{8}
	\widehat{\Lambda}(s,t)=n^{-1}\sum_{i=1}^n \hat{\nu}_i(s) \hat{\nu}_i(t).
\end{align}

Next, we estimate the variance function of the measurement error $e_i(s)$. Denote $e_i^*(s) = y_i^{*}(s) - \hat{\nu}_i(s)$, and ${\bm e}^*_i = ({e}^*_i(s_1), \cdots, {e}^*_i(s_M))^\mathrm{T}$. 
Suppose that ${\mathcal{E}}(s,s) \in \mathcal{H}$, then $\widehat{\mathcal{E}}(s,s) =  {\bm K}_s^\mathrm{T}\hat{\bm g}$, where $\hat{\bm g}$ is obtained by
\begin{align*}
	\hat{\bm g}
	&=\mathop{\arg\min}\limits_{{\bm g}} \left\{\frac{1}{nM}\sum_{i=1}^n[{\bm e}_i^{*2} - {\bm K}{\bm g}]^\mathrm{T}[{\bm e}_i^{*2} - {\bm K}{\bm g}] + \lambda {\bm g}^\mathrm{T} {\bm K}{\bm g}\right\}\\
	&=({\bm K} + \lambda M {\bm I})^{-1} \left\{n^{-1}\sum_{i=1}^n {\bm e}_i^{*2}\right\}.
\end{align*} 
Combing with (\ref{8}), the RKHS estimator of the covariance matrix is $\widehat{\bm \Phi} =(\widehat{\Lambda}(s_k,s_l)+\widehat{\mathcal{E}}(s_k,s_k))_{k,l=1,\cdots,M}$.

Given the weight $\widehat{\bm \Phi}^{-1}$, we iterate steps in Section \ref{sec2.1} with replacing the loss function in (\ref{5}) by a weighted loss function. The estimation procedure for the weighted estimate is summarized in the Algorithm C1 in the supplementary material. We denote the weighted estimate as $(\breve{\bm \theta},\breve{\bm \gamma})$.

\begin{remark}
	\rm{The B-spline methods and FPCA methods have been employed to approximate $\nu_i(s)$ and $e_i(s)$, see \cite{lijialiang2017, liu2021}.
		In contrast to these existing methods, the proposed RKHS estimator is implemented without finite truncation.} 	
\end{remark}

\section{Inference Procedure}\label{sec3}
Subgroup testing is essential to avoid false positive outcomes, whereas existing methods focus on the inference of non-dynamic subgroup effects. In this section, we consider the test for possible dynamic subgroup effects as follows:
\begin{gather*}
	H_{0}: {\bm \delta}(s)= 0,  \text{ for all } s \in \mathcal{S} \quad \text{ versus } \quad H_{1}: {\bm \delta}(s) \neq 0 , \text{ for some } s \in \mathcal{S}.
\end{gather*}
Without loss of generality, we assume that $\mathcal{S}=[0, 1]$. 

\subsection{Test Statistic}
Denote $\{{\bm A}_i(s)=(Y_i(s),{\bm X}_i^\mathrm{T},\tilde{\bm X}_i^\mathrm{T}, {\bm Z}_i^\mathrm{T})^\mathrm{T}, i=1,\cdots, n\}$ as the $n$ copies of process ${\bm A}(s)=(Y(s),{\bm X}^\mathrm{T},\tilde{\bm X}^\mathrm{T},{\bm Z}^\mathrm{T})^\mathrm{T}$.
We begin with formulating a score estimating equation under the null hypothesis. 
By calculating the Fr\'{e}chet derivatives of loss function $\mathcal{L}_{nM}(\bm \eta)$ with respect to ${\bm \delta}(s)$ in the $L_2$ space, the estimating equation is
$$ \sum_{i=1}^{n} [Y_{i}(s)-{\bm X}_{i}^{\mathrm{T}}{\bm \beta}(s)- \tilde{\bm X}_i^\mathrm{T}{\bm \delta}(s) I(Z_{1i}+{\bm Z}_{2i}^\mathrm{T} {\bm \gamma} > 0)]\tilde{\bm X}_iI(Z_{1i}+{\bm Z}_{2i}^\mathrm{T} {\bm \gamma} > 0)=0.$$
For a given ${\bm \gamma}$ and any $s \in [0,1]$, we consider the following score function with respect to ${\bm \delta}(s)$ under $H_0$,
\begin{align}\label{10}
	\Psi_{1n}(\tilde{\bm \beta}(s),0, {\bm \gamma})
	=\frac{1}{n} \sum_{i=1}^{n}[Y_{i}(s)-{\bm X}_{i}^{\mathrm{T}} \tilde{\bm \beta}(s)]\tilde{\bm X}_{i}I(Z_{1i}+{\bm Z}_{2i}^\mathrm{T} {\bm \gamma} > 0),
\end{align}
where $\tilde{\bm \beta}(s)$ is an estimator of ${\bm \beta}(s)$ under the null according to
\begin{align*}
	\tilde{\bm \beta}(s)=\mathop{\arg\min}\limits_{\bm \beta} \left\{\frac{1}{nM} \sum_{i=1}^{n} \sum_{m=1}^{M}\left[Y_{i}\left(s_m\right)-{\bm X}_{i}^{\mathrm{T}} {\bm \beta}(s_m)\right]^{2} + {\lambda} J(\bm \beta, \bm \beta)\right\}.
\end{align*}
For notational simplicity, denote $\psi_1({\bm A}_i(s),{\bm \beta}(s),0, \bm \gamma)=[Y_{i}(s)-{\bm X}_{i}^{\mathrm{T}} {\bm \beta}(s)]\tilde{\bm X}_{i}I(Z_{1i}+{\bm Z}_{2i}^\mathrm{T} {\bm \gamma} > 0)$ and $\psi_2({\bm A}_i(s),\bm \beta (s))={m}^{-1}\sum_{m=1}^{M}[Y_i(s_m)-{\bm X}_i^\mathrm{T} {\bm \beta}(s_m)]{\bm X}_iK(s_m,s)$.

Note that the grouping parameter $\bm \gamma$ is not identified under $H_0$ and can be regarded as a nuisance
parameter under $H_1$, we propose the supremum of squared score test statistic over the parametric space of $\bm \gamma$ based on the score function (\ref{10}), that is,
\begin{align*}
	T_n=\sup_{\bm \gamma} \int_{0}^1 n\Psi_{1n}(\tilde{\bm \beta}(s),0,\bm \gamma)^\mathrm{T} \widehat{\bm V}_S(s,\bm \gamma)^{-1} \Psi_{1n}(\tilde{\bm \beta}(s),0,\bm \gamma)  d s,
\end{align*}
where $\widehat{\bm V}_S(s,\bm \gamma)$ is a consistent estimate of the asymptotic variance of ${n}^{1/2}\Psi_{1n}(\tilde{\bm \beta}(s),0,\bm \gamma)$ and the definition of $\widehat{\bm V}_S(s,\bm \gamma)$ is given in Section \ref{sec3.2}.
\begin{remark}
	\rm{\cite{songrui2017} and  \cite{huang2020} proposed a similar test statistic for subgroup testing in binary data. Our test statistic can be seen as an extension of their approach, adapted for functional data.}
\end{remark}

\subsection{Resampling Method}\label{sec3.2}
In this subsection, we propose a general resampling method applicable to functional data and calculate the $p$-value of the proposed test statistic $T_n$.
Define
\begin{align*}
	\hat{\psi}_{*}({\bm A}_i,\tilde{\bm \beta},0,{\bm \gamma}) =  \psi_1({\bm A}_i,\tilde{\bm \beta},0,{\bm \gamma}) - \hat{D}({\bm \gamma}) \hat{J}^{-1}\psi_2({\bm A}_i,\tilde{\bm \beta}),
\end{align*}
where $\hat{D}({\bm \gamma})$ and $\hat{J} $ are consistent estimators of $D({\bm \gamma}) = \partial E\psi_1({\bm A}_i,{\bm \beta}_0,0,{\bm \gamma})/\partial {\bm \beta}^\mathrm{T}$ and $J = \partial E\psi_2({\bm A}_i,\bm \beta_0)/\partial {\bm \beta}^\mathrm{T}$, respectively.
We now consider the following perturbed test statistic
\begin{align}\label{eq3.2}
	T_n^{*}=\sup_{\gamma} \int_{0}^1 n\Psi_{1n}^{*}(\tilde{\bm \beta}(s),0,\bm \gamma)^\mathrm{T} \widehat{\bm V}_S(s, \bm \gamma)^{-1} \Psi_{1n}^{*}(\tilde{\bm \beta}(s),0,\bm \gamma)  d s,
\end{align}
where $\Psi_{1n}^{*}(\tilde{\bm \beta}(s),0,\bm \gamma) = {n}^{-1} \sum_{i=1}^{n} \xi_i \hat{\psi}_{*}({\bm A}_i(s),\tilde{\bm \beta}(s),0,\bm \gamma)$, $\widehat{\bm V}_S(s,\bm \gamma) =n^{-1}\sum_{i=1}^n \hat{\psi}_{*}({\bm A}_i(s),\tilde{\bm \beta}(s),0,{\bm \gamma})^{\otimes 2}$ with ${\bm v}^{\otimes 2} = {\bm v}{\bm v}^\mathrm{T}$ for any vector ${\bm v}$,
and $\{\xi_1, \cdots, \xi_n\}$ are independent and identically distributed from standard normal distribution $\mathcal{N}(0,1)$.

Since it is difficult to obtain the exact form of the test statistic $T_n^*$,
we follow \cite{huang2020} to find the maximum at a sequence of threshold values $\{{\bm \gamma}_1, \cdots, {\bm \gamma}_Q\}$.
That is, we compute $$T_{n,\max}^* = \max\{T_n^*({\bm \gamma}_1), \cdots, T_n^*({\bm \gamma}_Q)\},$$
where 
$T_n^*(\bm \gamma)=M^{-1}\sum_{m=1}^M n\Psi_{1n}^*(\tilde{\bm \beta}(s_m),0,\bm \gamma)^\mathrm{T} \widehat{\bm V}_S(s_m,\bm \gamma)^{-1} \Psi_{1n}^*(\tilde{\bm \beta}(s_m),0,\bm \gamma).$
The resampling procedure for calculating the $p$-value of $T_{n,\max}^*$ is presented in Algorithm \ref{alg3}.

\begin{algorithm}[H]
	\caption{Calculate the $p$-value of $T_{n,\max}^*$ in Section \ref{sec3.2}}
	\label{alg3}
	\begin{algorithmic}[1]
		\STATE Draw independent random samples $\{\xi_i^{b}, i=1,\cdots n, b=1,\cdots, B \}$ from $\mathcal{N}(0,1)$.
		\STATE Compute $\Psi_{1n}^{*b}(\tilde{\bm \beta}(s_m),0,{\bm \gamma}_j)$ defined in (\ref{eq3.2}) with $\{\xi_i^{b},i=1,\cdots n\}$,
		and calculate $\widehat{\bm V}_S(s_m,\bm \gamma_j)$ for each ${\bm \gamma}_j$ and $s_m$.
		\STATE Calculate $T_{n}^{*b}(s_m, {\bm \gamma}_j)=n \Psi_{1n}^{*b}(\tilde{\bm \beta}(s_m),0,{\bm \gamma}_j)^\mathrm{T} \widehat{\bm V}_S(s_m,{\bm \gamma}_j)^{-1} \Psi_{1n}^{*b}(\tilde{\bm \beta}(s_m),0,{\bm \gamma}_j)$. 
		\STATE Compute $T_n^{*b}({\bm \gamma}_j) = M^{-1}\sum_{m=1}^M T_{n}^{*b}(s_m,{\bm \gamma}_j)$ with $j=1,\cdots, Q$, and obtain their maximum value $T_{n,\max}^{*b}=\max\{T_n^{*b}({\bm \gamma}_1),\cdots, T_n^{*b}({\bm \gamma}_Q)\}$ for $b=1,\cdots,B$.
		\STATE Calculate the $p$-value by $\#\{T_{n,\max}^{*b} > T_{n,\max}^{*}\}/B$.
	\end{algorithmic}
\end{algorithm}

\section{Asymptotic Properties}\label{sec4}
We, herein, study the asymptotic properties of the estimators and the test statistics proposed in Section \ref{sec2} and Section \ref{sec3}. Our theoretical results hold when both $n \rightarrow \infty$ and $M \rightarrow \infty$.

\subsection{Subgroup Identification Theory}
Before stating the theory results, we first introduce some notation and regularity assumptions. 
Let $S_{nM,{\lambda}}^{\bm \theta}({\bm \eta};h)$ be the first Fr\'{e}chet derivative of $\mathcal{L}_{nM, {\lambda}}({\bm \eta}; h)$ with respect to ${\bm \theta}$ and $\mathcal{D}$ be the Fr\'{e}chet derivative operator.
Denote ${\bm W}_{i,\bm \gamma} = ({\bm X}_{i}^{\mathrm{T}}, \tilde{\bm X}_{i}^{\mathrm{T}}G_h(Z_{1i}+{\bm Z}_{2i}^\mathrm{T} \bm \gamma))^\mathrm{T}$ with a given ${\bm \gamma}$.
Then for any ${\bm \theta}_1, {\bm \theta}_2 \in \mathcal{H}^{p+d}$, 
\begin{align}\label{14}
	S_{nM,\lambda}^{\bm \theta}({\bm \eta};h) {\bm  \theta_1}  &=-\frac{1}{nM} \sum_{i=1}^{n}\sum_{m=1}^M [Y_{i}(s_m)-{\bm W}_{i,\bm \gamma}^{\mathrm{T}} {\bm \theta}(s_m)] {\bm W}_{i,\bm \gamma}^{\mathrm{T}} {\bm \theta}_1(s_m) + \lambda J({\bm \theta}, {\bm \theta}_1),\\ 
	\mathcal{D}S^{\bm \theta}_{nM,\lambda}({\bm \eta};h) {\bm  \theta_1} { \bm \theta_2} & = \frac{1}{nM} \sum_{i=1}^{n} \sum_{m=1}^{M}{\bm \theta_1}(s_{m})^{\mathrm{T}}{\bm W}_{i,\bm \gamma}{\bm W}_{i,\bm \gamma}^{\mathrm{T}} { \bm  \theta_2}(s_m) + \lambda J({\bm \theta}_1, {\bm \theta}_2), \nonumber 
\end{align}
where $J(\bm \theta, \bm \theta_1) = \left\langle {\bm \theta},  {\bm \theta}_1 \right\rangle_{K}$. Let $S^{\bm \theta}_{\lambda}({\bm \eta};h)=E\{S_{nM,\lambda}^{\bm \theta}({\bm \eta};h)\}$, it follows from (\ref{14}) that
\begin{align}\label{15}
	\mathcal{D}S^{\bm \theta}_{\lambda}({\bm \eta};h){\bm \theta}_1 {\bm \theta}_2 = \int_{\mathcal{S}} {\bm \theta}_1(s)^\mathrm{T} E({\bm W}_{\bm \gamma}{\bm W}_{\bm \gamma}^{\mathrm{T}}) {\bm \theta}_2(s) \pi(s) ds + \lambda J({\bm \theta}_1, {\bm \theta}_2),
\end{align}
where $\pi(s)$ is the density function of $s_m$. Motivated by \cite{cheng2015} and \cite{shang2015}, we define the inner product $\langle \cdot, \cdot \rangle_{\lambda} : \mathcal{H}^{p+d} \times \mathcal{H}^{p+d} \rightarrow \mathbb{R}$ by 
\begin{align}\label{6}
	\langle {\bm \theta}_1, {\bm \theta}_2 \rangle_{\lambda} =  V({\bm \theta}_1, {\bm \theta}_2) +  \lambda J({\bm \theta}_1,  {\bm \theta}_2),
\end{align}
where $V({\bm \theta}_1, {\bm \theta}_2) = \int_{\mathcal S} {\bm \theta}_1(s)^\mathrm{T} E({\bm W}_{\bm \gamma_0}{\bm W}_{\bm \gamma_0}^{\mathrm{T}}) {\bm \theta}_2(s)\pi(s)ds$, and $\bm \gamma_0$ is the true value of $\bm \gamma$.
The corresponding norm for the inner product (\ref{6}) is denoted as $\|\cdot\|_{\lambda}$.
According to (\ref{15}) and (\ref{6}), we further have 
\begin{align*}
	\mathcal{D}S^{\bm \theta}_{\lambda}({\bm \eta}_0;h){\bm \theta}_1 {\bm \theta}_2 = \langle {\bm \theta}_1, {\bm \theta}_2 \rangle_{\lambda},
\end{align*}
which implies that $\mathcal{D}S^{\bm \theta}_{\lambda}({\bm \eta}_0;h) = id$, where $id$ is the identity operator in $\mathcal{H}^{p+d}$.

Let $\mathcal{R}_{\lambda}(s_1, s_2)$ be the reproducing kernel matrix of $\mathcal{H}^{p+d}$ endowed with norm $\|\cdot\|_{\lambda}$. That is, the kernel $\mathcal{R}_{\lambda}(\cdot, \cdot)$ has the reproducing property that for any ${\bm \theta} \in \mathcal{H}^{p+d}$, ${\bm c} \in \mathbb{R}^{p+d}$ and $s_1, s_2  \in \mathcal{S}$, $(\mathcal{R}_{\lambda, s_1}{\bm c})(s_2) = \mathcal{R}_{\lambda}(s_1,s_2){\bm c}$ and $\langle \mathcal{R}_{\lambda, s_1}{\bm c}, {\bm \theta} \rangle_{\lambda} = {\bm c}^\mathrm{T}{\bm \theta}(s_1)$. More properties about the vector-valued RKHS have been provided by \cite{minh2016} and \cite{hao2022}. 
We assume that there exists a sequence of basis functions in the space $\mathcal{H}^{p+d}$ that diagonalizes the operators $V$ and $J$ defined in (\ref{6}) simultaneously. Such a diagonalization assumption is typical in the literature, see \cite{caiyuan2011,cheng2015,shang2015}.

\begin{assumption}
	\rm{There exists a sequence of vector valued functions ${\bm h}_l \in \mathcal{H}^{p+d}$ such that $\sup_l\sup_{s}|{\bm h}_l(s) | < \infty$, and
		$V({\bm h}_i, {\bm h}_l) = \delta_{il}$, $J({\bm h}_i, {\bm h}_l)=\rho_l^{-1}\delta_{il}$, where $\delta_{il}$ is the Kronecker delta, and $\rho_{l} \asymp l^{-2\alpha}$. Furthermore, any $\bm \theta \in \mathcal{H}^{p+d}$ admits the expansion $\bm \theta = \sum_{l=1}^{\infty}V(\bm \theta, {\bm h}_l){\bm h}_l$ with convergence in $\mathcal{H}^{p+d}$ under the norm $\|\cdot\|_{\lambda}$.}
\end{assumption}

\begin{remark}
	\rm{Under Assumption 1, then $\|{\bm \theta}\|_{\lambda}^2 = \sum\limits_{l=1}^{\infty}V({\bm \theta}, {\bf h}_l)^2(1+\lambda \rho_l^{-1})$, and $\mathcal{R}_{\lambda,s}(\cdot) = \sum_{l=1}^{\infty} {{\bm h}_l(s){\bm h}_l(\cdot)^\mathrm{T}}/({1+\lambda \rho_l^{-1}})$.}
\end{remark}

We next state regularity assumptions to obtain the consistency results.  To facilitate the expression of asymptotic results, we perform linear transformations on the variables.
Denote ${\bm T}_i$ as the variables including ${\bm X}_i, {\bm Z}_i$, and ${\bm T}_{2i}$ as the variables in ${\bm T}_i$ excluding $Z_{1i}$.
Let $q_i = Z_{1i} + {\bm Z}_{2i}^\mathrm{T} {\bm \gamma}_0$ and ${\bm W}_{i}^* = ({\bm X}_{i}^{\mathrm{T}}, \tilde{\bm X}_{i}^{\mathrm{T}}I(q_i > 0))^\mathrm{T}$. 
There is a one-to-one relation between $(q_i, {\bm T}_{2i}^\mathrm{T})^\mathrm{T}$ and ${\bm T}_{i}$, which represents that there exists $(\dot{\delta}_{10}(s), \dot{\bm \delta}_{20}(s)^\mathrm{T})^\mathrm{T}$ such that $\tilde{\bm X}_i^\mathrm{T} {\bm \delta}_0(s) = q_i\dot{\delta}_{10}(s)+{\bm T}_{2i}^\mathrm{T}\dot{\bm \delta}_{20}(s)$.
Denote $f_{q|{\bm T}_2}(q|{\bm T}_2)$ as the density of $q$ conditional on ${\bm T}_2$,
and define $f^{(i)}_{q|{\bm T}_2}(q|{\bm T}_2) = \partial^if_{q|{\bm T}_2}(q|{\bm T}_2)/\partial q^i$ whenever the derivative exists. Similar transformation has been made in \cite{seolinton2005} and \cite{zhangyy2021}.

\begin{assumption}\rm{
		\begin{enumerate}[label=(\roman*)]
			\item The covariates $\bm T_i$ are independent and identically distributed, and $\sup_i\|\bm T_i\|_2 < \infty$ almost surely.  
			\item The true coefficients $\bm \gamma_0$ are in the interior of compact subspaces $\Upsilon$, and $0<P(Z_{1i}+{\bm Z}_{2i}^\mathrm{T}{\bm \gamma}>0)<1$ for any $\bm \gamma \in \Upsilon$.
			\item For almost every ${\bm Z}_{2i}$, the density of $Z_{1i}$ conditional on ${\bm Z}_{2i}$ is everywhere positive.
			\item The minimum eigenvalue of $E\{{\bm W}^{*\otimes 2} | {\bm Z}\}$ is bounded away from zero uniformly over ${\bm Z}$.
			\item For all $m=1,\cdots,M$, $\sup_{s_m}E[|e_{ij}(s_m)|^b] < \infty$ for some $b>4$ and all grid points $s_m$.
			\item The functional classes $\{{\nu}(s): s \in [0,1]\}$ $\{{\nu}(s){\nu}(t): (s,t) \in [0,1]^2\}$ are Donsker classes, and $E[\sup_{s \in [0,1]}|\nu(s)|^a] < \infty$ for some $a > 2$. All components of $\Lambda(s, t)$ have continuous second-order partial derivatives with respect to $(s,t) \in [0,1]^2$ and $\inf_{s\in [0,1]}\Lambda(s, s) > 0$.	
			\item The grid points $\mathcal{S} = \{s_m, m = 1, \cdots, M\}$ are independently and identically distributed with density function $\pi(s)$. For all $s \in [0,1]$, $\pi(s)>0$ and $\pi(s)$ has continuous second-order derivative.
	\end{enumerate}}
\end{assumption}
Assumption 2 (\romannumeral1) is assumed for theoretical convenience, the boundedness of $\bm T_i$ is not essential. 
Assumption 2 (\romannumeral2) is satisfied for many commonly used change-plane models, see \cite{songrui2017}.
Assumption 2 (\romannumeral3) is used to establish the asymptotic equivalence between the smoothed and nonsmoothed objective functions as $h \rightarrow 0 $. 
Assumption 2 (\romannumeral4) ensures the regression coefficient identification in change-plane or threshold models. Assumption 2 (\romannumeral2)-(\romannumeral4) are commonly assumed in the literature \citep{zhangyy2021,wangxiao2020}. 
Assumption 2 (\romannumeral5)-(\romannumeral7) are standard regularity conditions in the functional data analysis literature \citep{zhu2012,lijialiang2017}. 


\begin{theorem}\label{th1}
	Suppose that Assumptions 1 and 2 hold, if $h \rightarrow 0$,
	then $\hat{\bm \eta} \rightarrow {\bm \eta}$ in probability.
\end{theorem}

\begin{theorem}\label{th11}
	Suppose that Assumptions 1 and 2 hold, $h\lambda^{-1/(2\alpha)}=o(1)$, $nh\lambda^{1-1/(2\alpha)}=o(1)$ and $M^{-1}\lambda^{-1/\alpha}h =o(1)$, then
	$\|\hat{\bm \theta} - {\bm \theta_0}\|_{\lambda} = O_p((nM\lambda^{1/(2\alpha)})^{-1/2} + n^{-1/2} + \lambda^{1/2}).$
\end{theorem}

When the tuning parameter $\lambda=O((nM)^{-2\alpha/(2\alpha+1)})$, the convergence rate leads to $O_p(n^{-1/2} + (nM)^{-\alpha/(2\alpha+1)})$, which is the optimal convergence rate for the independent design in functional regression model \citep{caiyuan2011,wangxiao2020}.
The sampling frequency affects the convergence rate.
For a dense sampling like $M \gg n^{1/(2\alpha)}$, the convergence rate is dominated by $n^{-1/2}$, which is caused by the high spatial dependency between the observed responses. While for a sparse sampling like $M \ll n^{1/(2\alpha)}$, the convergence rates are dominated by $(nM)^{-\alpha/(2\alpha+1)}$. In this situation, the result is consistent with the optimal nonparametric convergence rate as if all the $nM$ observations are independently observed.
Based on the consistency of $(\hat{\bm \theta}, \hat{\bm \gamma})$, we next derive their asymptotic distribution.
Our theoretical analysis also adopts the following additional regularity conditions to establish the asymptotic normality.

\begin{assumption}\rm{
		\begin{enumerate}[label=(\roman*)]
			\item The smooth function $G(\cdot)$ is twice differentiable and $G(s) + G(-s)=1$. $G'(\cdot)$ is symmetric around zero. Both $|G'(w)|$ and $|G''(w)|$ are uniformly bounded over $w$. $\int |G'(w)| dw < \infty$ and $\int |G''(w)| dw < \infty$. Furthermore, $\int w \{G(w)-I(w>0)\} dw < \infty$, $\int \{I(w>0)-G(w)\}^2dw < \infty$ and $\int \{I(w>0)-G(w)\}^2G'(w)^2 dw < \infty$.
			
			\item $f^{(k)}_{q|{\bm T}_2}(q|{\bm T}_2)$ is a continuous function and $|f^{(k)}_{q|{\bm T}_2}(q|{\bm T}_2)|<C$ uniformly over $(q, {\bm T}_2)$ for each integer $k$ with $0 \leq k \leq k'$(defined later).
			
			\item For each integer $k$ with $1\leq k \leq k'$, $\int w^{k-1}\{G(w)-I(w>0)\}G'(w)dw=0$ and $\int w^{k'}\{G(w)-I(w>0)\}G'(w)dw \neq 0$.

			\item $nh^2 \rightarrow \infty$, $nh^3 \rightarrow 0$ if $k'=1$ and $nh^4 \rightarrow 0$ if $k' > 1$. 
			
			\item The eigenvalues of ${\bm V}^{\bm \gamma}$ (defined later) are bounded below and above by some positive constants $c_1$ and $1/c_1$, respectively. 
			The eigenvalues of $\mathcal{Q}^{\bm \gamma}$ (defined later) are bounded below and above by some positive constants $c_2$ and $1/c_2$, respectively. 
	\end{enumerate}}
\end{assumption}
Assumption 3 (\romannumeral1)-(\romannumeral5) are regular conditions in change-plane models, see \cite{seolinton2005} and \cite{zhangyy2021}.
Assumption 3 (\romannumeral5) ensures that ${\bm V}^{\bm \gamma}$, $\mathcal{Q}^{\bm \gamma}$ are positive definite matrix.

\begin{theorem}\label{th2}
	Suppose that Assumptions in Theorem \ref{th11} and Assumption 3 hold, ${n}^{1/2}\lambda = o(1)$, $\sum_l \rho_l^{-2} V({\bm \theta}_0, {\bm h}_l)^2 < \infty$ and $\lambda^{-1/(2\alpha)}M^{-1} = o(1)$.
	Then
	(i) for any $s \in [0, 1]$, ${{n}^{1/2}}{\bm V}^{\bm \theta}(s)^{-1/2}(\hat{\bm \theta}(s) - {\bm \theta_0}(s)) \rightarrow \mathcal{N}(0, {\bm I}_{p+d})$ in distribution,  ${n}^{1/2}{h}^{-1/2}\{[\mathcal{Q}^{\bm \gamma}]^{-1}{\bm V}^{\bm \gamma}[\mathcal{Q}^{\bm \gamma}]^{-1}\}^{-1/2}\\(\hat{\bm \gamma}-{\bm \gamma}_0) \rightarrow \mathcal{N}(0, {\bm I}_q)$ in distribution,	
	and the two are asymptotically independent, where ${\bm V}^{\bm \theta}(s) = \mathcal{\bm V}^{\bm \theta}(s,s)$,
	\begin{align*}
		\mathcal{\bm V}^{\bm \theta}(s,t) =& M^{-2}E\left\{[\mathcal{R}_{\lambda}(s_1,s),\cdots,\mathcal{R}_{\lambda}(s_M,s)][(I_M \otimes {\bm W}_{i}^*) \Lambda^{1/2}]^{\otimes 2} [\mathcal{R}_{\lambda}(s_1,t),\cdots,\mathcal{R}_{\lambda}(s_M,t)]^\mathrm{T} \right\}, 
	\end{align*}
	with $\Lambda = \{\Lambda(s_m, s_{m'})\}_{m,m'=1,\cdots,M}$, and
	\begin{align*}
		{\bm V}^{\bm \gamma} = &\int G'(w)^2dw E_{{\bm Z}_2, \bm{T}_2}\left\{\int\int\bm{T}_2^\mathrm{T} \dot{\bm \delta}_{20}(s) \Lambda(s,t)\bm{T}_2^\mathrm{T} \dot{\bm \delta}_{20}(t)dsdt {\bm Z}_{2}{\bm Z}_{2}^\mathrm{T} f_{q|\bm{T}_2}(0|\bm{T}_2)\pi(s)\pi(t)\right\},\\
		\mathcal{Q}^{\bm \gamma}=&G'(0) E_{{\bm Z}_2, \bm{T}_2}\left\{ \int [{\bm T}_2^\mathrm{T} \dot{\bm \delta}_{20}(s)]^2 ds {\bm Z}_{2}{\bm Z}_{2}^\mathrm{T} f_{q|{\bm T}_2}(0|{\bm T}_2)\pi(s)\right\}.
	\end{align*}
	(ii) if $\lim_{\lambda \rightarrow 0}\lambda^{1/(2\alpha)}\sum_l \{[{\bm h_l(t)}({\bm h}_l(s_1)-{\bm h}_l(s_2))^\mathrm{T}]^{\otimes 2}/[(1+\lambda\rho_l^{-1})^2]\} \leq c_0|s_1-s_2|^{2\vartheta}$ for some nonnegative constants $c_0, \vartheta$, then $\{{n}^{1/2}(\hat{\bm \theta}(s) - {\bm \theta_0}(s)) : s \in [0,1]\}$ converges weakly to a mean-zero $(p+d)$ dimensional Gaussian process with covariance matrix $\mathcal{\bm V}^{\bm \theta}(s,t)$.
\end{theorem}

As observed in the change-plane literature for scalar data, the asymptotic results of coefficient estimators $\hat{\bm \theta}(s)$ and change plane parameter $\hat{\bm \gamma}$ have different convergence rates. It follows from Theorem \ref{th2} that the converges rate of $\hat{\bm \gamma}$ is $h^{1/2}n^{-1/2}$, while the convergence rate of $\hat{\bm \theta}(s)$ is $(n\lambda^{{1}/({2\alpha})})^{-1/2}$. Therefore, the convergence rate of $\hat{\bm \gamma}$ is faster than $\hat{\bm \theta}(s)$, which is also consistent with the results on scalar data, see more details in \cite{seolinton2005} and \cite{zhangyy2021}. 
We next study the consistency of estimated covariance function $\widehat{\Lambda}(s, s')$ and $\widehat{\mathcal{E}}(s, s')$.

\begin{theorem}\label{th3}
	Suppose that Assumptions in Theorem \ref{th2} hold. Then
	(i)$\sup\limits_{s, s' \in \mathcal{S}} |\widehat{\Lambda}(s, s')-{\Lambda}(s, s')| = o_p(1)$;
	(ii)$\sup\limits_{s \in \mathcal{S}}
	|\widehat{\mathcal{E}}(s, s)-{\mathcal{E}}(s, s)| = o_p(1).$
\end{theorem}

Combining the results of Theorem \ref{th2} and Theorem \ref{th3}, the weighted estimates of function parameter and grouping parameter are also consistent.
Theorem \ref{th4} provides the limit distribution of weighted estimators $(\breve{\bm \theta}, \breve{\bm \gamma})$.

\begin{theorem}\label{th4}
	Suppose that Assumptions in Theorem \ref{th2} hold. Then
	\begin{enumerate}
		\item[(i)]for any $s \in [0, 1]$, ${n^{1/2}}\breve{{\bm V}}^{\bm \theta}(s)^{-1/2}(\breve{\bm \theta}(s) - {\bm \theta_0}(s)) \rightarrow \mathcal{N}(0, {\bm I}_{p+d})$ in distribution,
		where $\breve{\bm V}^{\bm \theta}(s) = \breve{\mathcal{\bm V}}^{\bm \theta}(s,s)$,
		\begin{align*}
			\breve{\mathcal{\bm V}}^{\bm \theta}(s,t)= M^{-2}E\left\{[\mathcal{R}_{\lambda}(s_1,s), \cdots, \mathcal{R}_{\lambda}(s_M,s)][(I_M \otimes {\bm W}_{i}^*) {\bm \Phi}^{-1/2}]^{\otimes 2} [\mathcal{R}_{\lambda}(s_1,t), \cdots, \mathcal{R}_{\lambda}(s_M,t)]^\mathrm{T}\right\},	 
		\end{align*}
		and $n^{1/2}h^{-1/2}\{\{\breve{\mathcal{Q}}^{\bm \gamma}\}^{-1}\breve{\bm V}^{\bm \gamma}\{\breve{\mathcal{Q}}^{\bm \gamma}\}^{-1}\}^{-1/2}(\breve{\bm \gamma}-{\bm \gamma}_0) \rightarrow \mathcal{N}(0, {\bm I}_q)$ in distribution, where
		\begin{align*}
			\breve{\bm V}^{\bm \gamma} = &\int G'(w)^2dw E_{{\bm Z}_2, \bm{T}_2}\left\{\int\int \bm{T}_2^\mathrm{T} \dot{\bm \delta}_{20}(s){\bm \Phi}^{-1}(s,t) \bm{T}_2^\mathrm{T} \dot{\bm \delta}_{20}(t) dsdt {\bm Z}_{2}{\bm Z}_{2}^\mathrm{T} f_{q|\bm{T}_2}(0|\bm{T}_2)\pi(s)\pi(t)\right\},\\
			\breve{\mathcal{Q}}^{\bm \gamma}=&G'(0) E_{{\bm Z}_2, \bm{T}_2}\left\{ \int\int {\bm T}_2^\mathrm{T} \dot{\bm \delta}_{20}(s) {\bm \Phi}^{-1}(s,t) {\bm T}_2^\mathrm{T} \dot{\bm \delta}_{20}(t)dsdt {\bm Z}_{2}{\bm Z}_{2}^\mathrm{T} f_{q|{\bm T}_2}(0|{\bm T}_2)\pi(s)\pi(t)\right\}.
		\end{align*}
		\item[(ii)]if $\lim_{\lambda \rightarrow 0}\lambda^{1/(2\alpha)}\sum_l \{[{\bm h_l(t)}({\bm h}_l(s_1)-{\bm h}_l(s_2))^\mathrm{T}]^{\otimes 2}/[(1+\lambda\rho_l^{-1})^2]\} \leq c_0|s_1-s_2|^{2\vartheta}$ for some nonnegative constants $c_0, \vartheta$, then $\{{n}^{1/2}(\breve{\bm \theta}(s) - {\bm \theta}_0(s)) : s \in [0,1]\}$ converges weakly to a $(p+d)$ dimensional mean-zero Gaussian process with covariance matrix $\breve{\mathcal{\bm V}}^{\bm \theta}(s,t)$.
	\end{enumerate}
\end{theorem}	

\subsection{Subgroup Testing Theory}
We first establish the limit distribution of the proposed test statistic $T_n$ under the null hypothesis in Theorem \ref{th6}(i) below. Next, we investigated the power performance of the proposed test statistic under the local alternative hypotheses where subgroups exist.
Theorem \ref{th6}(ii) below provides the asymptotic distribution of the test statistic $T_n$ under a sequence of local alternatives $H_{1n}: {\bm \delta}(s) = n^{-1/2}{\bm \tau}(s)$ with $n$ goes to infinity.

\begin{theorem}\label{th6}
	Suppose that Assumptions in Theorem \ref{th2} hold. Then
	\begin{enumerate}
		\item[(i)]under $H_0$, $T_n \rightarrow \sup_{\bm \gamma \in \Upsilon} \int_{0}^1 \mathbb{G}(s, \bm \gamma)^\mathrm{T} \mathbb{G}(s, \bm \gamma) d s$ in distribution,
		where $\{\mathbb{G}(s, \bm \gamma) : \bm \gamma \in \Upsilon\}$ is a Gaussian process with mean zero and covariance $$\varGamma(s, s', \bm \gamma_1, \bm \gamma_2) = {\bm V}(s,\bm \gamma_1)^{-1/2}E\{\psi_{*}({\bm A}_i(s),\bm \beta_{0}(s),0,\bm \gamma_1)\psi_{*}({\bm A}_i(s'),\bm \beta_{0}(s'),\bm \gamma_2)^\mathrm{T}\}{\bm V}(s',\bm \gamma_2)^{-1/2},$$
		for any $s, s' \in [0,1]$, $\bm \gamma_1, \bm \gamma_2 \in \Upsilon$,
		where ${\bm V}(s,\bm \gamma)=E\{\psi_{*}({\bm A}_i(s),\bm \beta_{0}(s),0,\bm \gamma)^{\otimes 2}\}$ and $\psi_{*}({\bm A}_i(s),\bm \beta_{0}(s),0,\bm \gamma) = \psi_{1}({\bm A}_i(s),\bm \beta_{0}(s),0,\bm \gamma)-D(s,\bm \gamma)J(s)^{-1}\psi_{2}({\bm A}_i(s),\bm \beta_{0}(s))$ with $D(s,{\bm \gamma})$ and $J(s)$ defined in Section \ref{sec3.2}.
		\item[(ii)]under the local alternatives $H_{1n}$, the statistic $T_n \rightarrow \sup_{\bm \gamma \in \Upsilon} \int_{0}^1 \mathbb{G}_{\bm \tau}(s,\bm \gamma)^\mathrm{T} \mathbb{G}_{\bm \tau}(s,\bm \gamma) d s$ in distribution,
		where $\{\mathbb{G}_{\bm \tau}(s, \bm \gamma) : \bm \gamma \in \Upsilon\}$ is a Gaussian process with mean function $$\mu(s,\bm \gamma)={\bm V}(s,\bm \gamma)^{-1/2}E\left\{\tilde{\bm X}_{i}\tilde{\bm X}_{i}^\mathrm{T} I(Z_{1i}+{\bm Z}_{2i}^\mathrm{T} {\bm \gamma}>0)I(Z_{1i}+{\bm Z}_{2i}^\mathrm{T} {\bm \gamma}_0 >0)\right\} {\bm \tau}(s),$$ and covariance function $\varGamma(s,s',\bm \gamma_1, \bm \gamma_2)$.
	\end{enumerate}
\end{theorem}

Theorem \ref{th6} establishes the limiting distribution of the test statistic $T_n$, which is a nontrivial extension of \cite{songrui2017} to functional responses.

\section{Simulation Studies}\label{sec5}
In this section, we present two simulation settings to demonstrate the performance of the proposed estimating and testing procedures.

\subsection{Subgroup Identification Model}\label{sec5.1}
This subsection is designed to evaluate the estimation method for the proposed functional change-plane model. We consider the following data-generating process
\begin{align*}
	Y_i(s_m) = {\bm X}_i^\mathrm{T} {\bm \beta}(s_m) + \tilde{\bm X}_i^\mathrm{T} {\bm \delta}(s_m)I(Z_{1i} + {\bm Z}_{2i}^\mathrm{T} {\bm \gamma} > 0) + \nu_{i}(s_m) + e_i(s_m),
\end{align*}
with $i=1, \cdots, n$ and $m=1, \cdots, M$. Assume that $\{s_m\} \sim U[0,1]$, ${\bm X}_i=(X_{1i},X_{2i},X_{3i})^\mathrm{T}$ is generated from multivariate normal distribution with mean zero, and the $(s,k)$th element of the covariance function is assumed to be $0.5^{|s-k|}$ for ${s,k=1,2,3}$. Moreover, $ \tilde{\bm X}_i=(X_{1i},X_{2i})^\mathrm{T}$, $Z_{1i} \sim \mathcal{N}(0,1)$ and ${\bm Z}_{2i} = (1, Z_{2i}^{*})^\mathrm{T}$ with $Z_{2i}^{*} \sim \mathcal{N}(1,1)$, $e_i(s) \sim \mathcal{N}(0, {0.1}^{1/2})$, $\nu_{i}(s)=\xi_{1i}\varsigma_1(s)+\xi_{2i}\varsigma_2(s)$, where $\varsigma_1(s) = {2}^{1/2}\sin(2\pi s)$, $\varsigma_2(s) = {2}^{1/2}\cos(2\pi s)$, and $\xi_{1i} \sim \mathcal{N}(0, 1)$, $\xi_{2i} \sim \mathcal{N}(0, 0.5)$. 
We set the grouping parameters ${\bm \gamma}=(\gamma_1, \gamma_2)^\mathrm{T}=(-1,1)^\mathrm{T}$ and the functional parameters ${\bm \beta}(s)=(\beta_1(s), \beta_2(s), \beta_3(s))^\mathrm{T}$ and ${\bm \delta}(s)=(\delta_1(s), \delta_2(s))^\mathrm{T}$ with
\begin{align*}
	&\beta_1(s) = (1-s)^3, \quad \beta_2(s)=\exp(-s^2), \quad \beta_3(s) = \sin(\pi s) + s^3, \\
	&\delta_1(s) = (1-s)^2 , \quad \delta_2(s)= \exp(-5s).
\end{align*}
A Gaussian kernel $K(s,s')=\exp\{-{|s-s'|^2}/({2\nu^2})\}$ with $\nu=0.2$ is used for $\mathcal{H}$. The nuisance parameters $\lambda=0.01$, and the bandwidth $h=n^{-1/2}\log(n)$.

To show the accuracy of identification of the subgroups, we define the accuracy rate as \cite{li2018}, that is,
$\text{Accuracy Rate} = 1- {n}^{-1}\sum_{i=1}^n  |I(Z_{1i} + {\bm Z}_{2i}^\mathrm{T} {\bm \gamma} > 0) - I(Z_{1i} + {\bm Z}_{2i}^\mathrm{T} \hat{\bm \gamma} > 0)|.$
To evaluate the performance of coefficient function estimators, we use the root-average squared errors (RASEs), which is defined as $\text{RASEs}(f)=\left[M^{-1}\sum_{m=1}^M (\hat{f}(s_m)-f(s_m))^2\right]^{1/2}.$
In the following Tables,  "LS" stands for the initial estimators in Section \ref{sec2.1}, and "WLS" stands for the weighted estimators improved by incorporating the spatial dependency in Section \ref{sec2.4}.

\begin{figure}[htbp]
	\centering
	\begin{minipage}[b]{0.48\linewidth}
		\includegraphics[width=1\linewidth, height=1\linewidth]{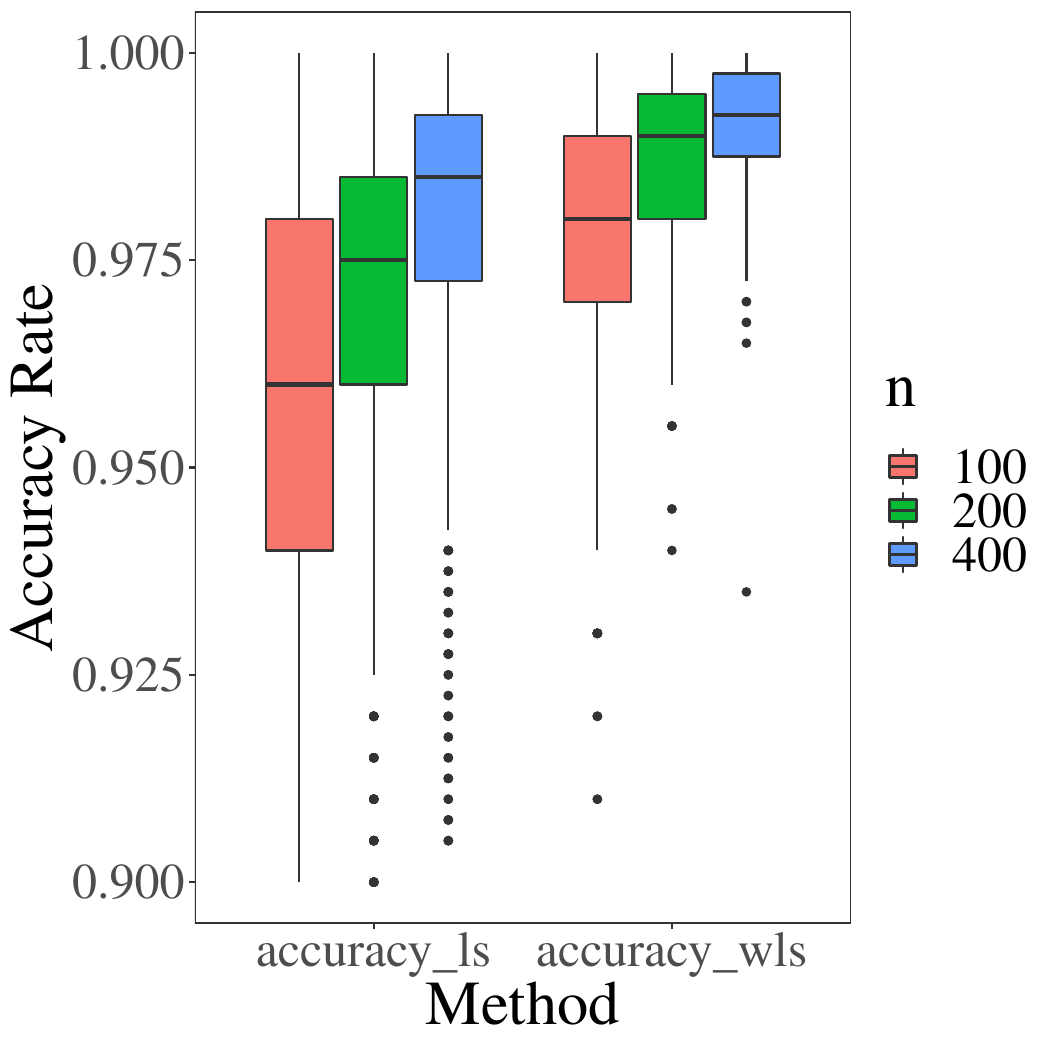}
	\end{minipage}
	\quad
	\begin{minipage}[b]{0.48\linewidth}
		\includegraphics[width=1\linewidth, height=1\linewidth]{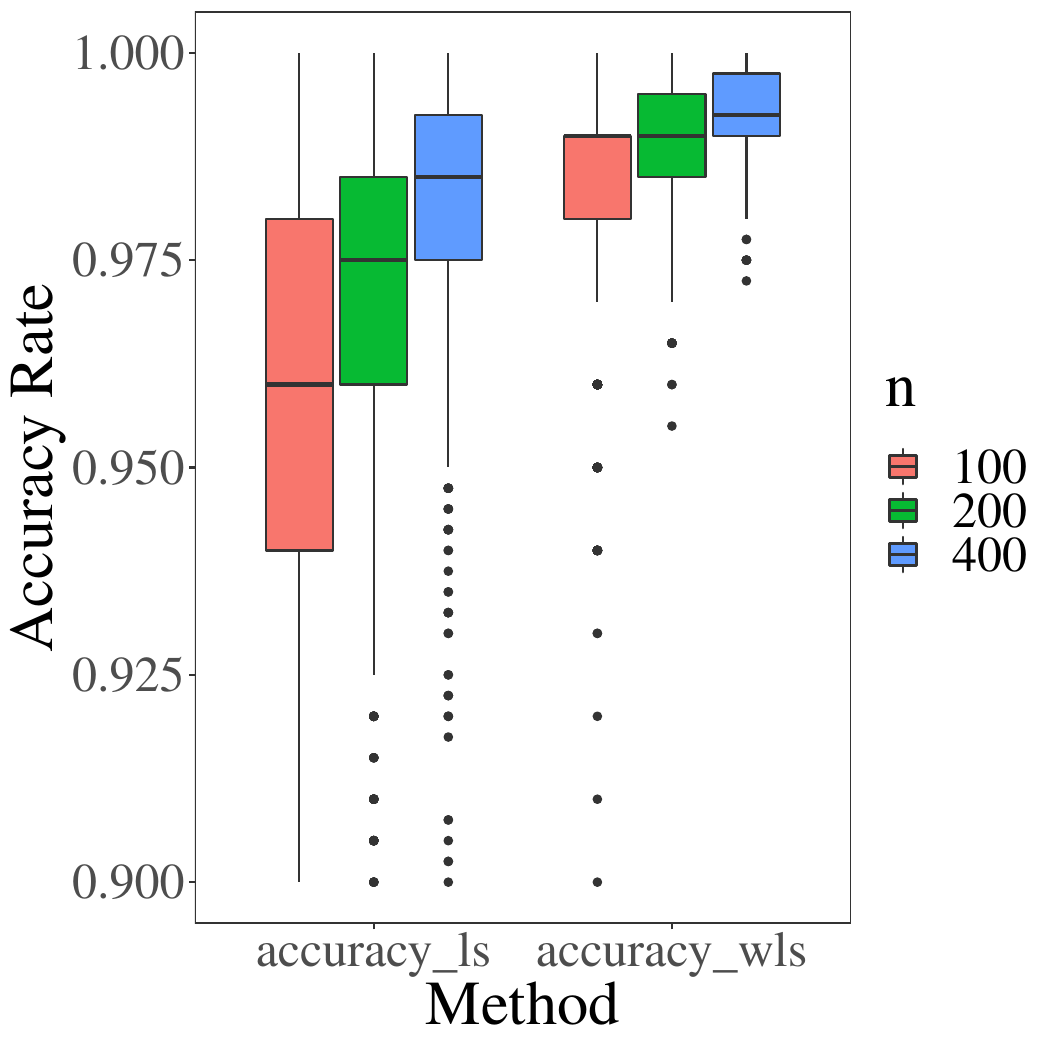}
	\end{minipage}
	\caption{The left panel shows the boxplot of accuracy rate with $M=10$; The right panel shows the boxplot of accuracy rate with $M=30$. \text{"accuracy\_ls"} represents the initial estimation method and \text{"accuracy\_wls"} represents the weighted estimation method.}
	\label{fig4}
\end{figure}

\begin{figure}[!htbp]
	\centering
	\subfigure[]
	{
		\includegraphics[width=0.31\linewidth,height=0.31\linewidth]{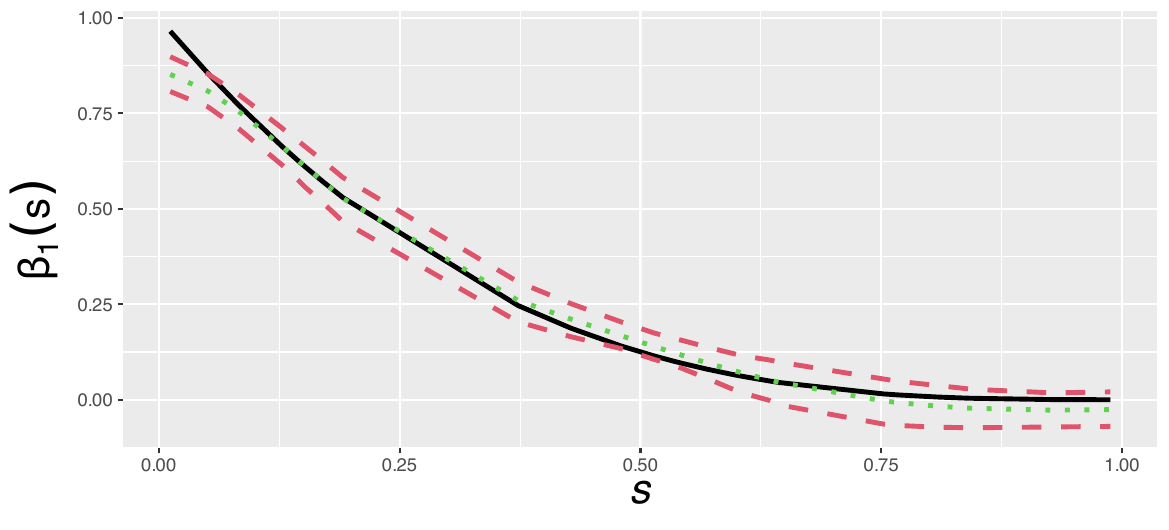}
	}
	\subfigure[]
	{
		\includegraphics[width=0.31\linewidth,height=0.31\linewidth]{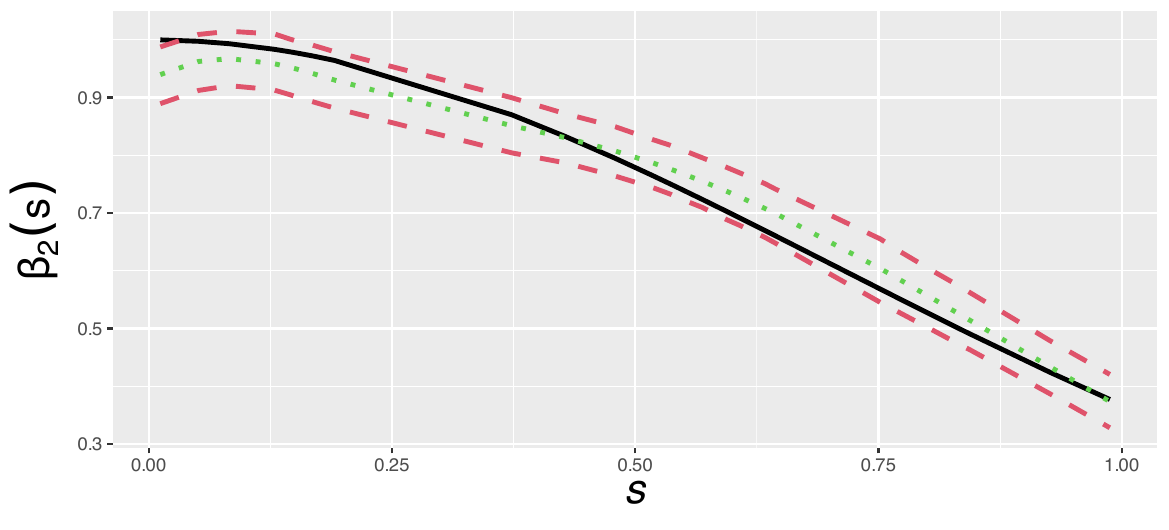}
	}
	\subfigure[]
	{
		\includegraphics[width=0.31\linewidth,height=0.31\linewidth]{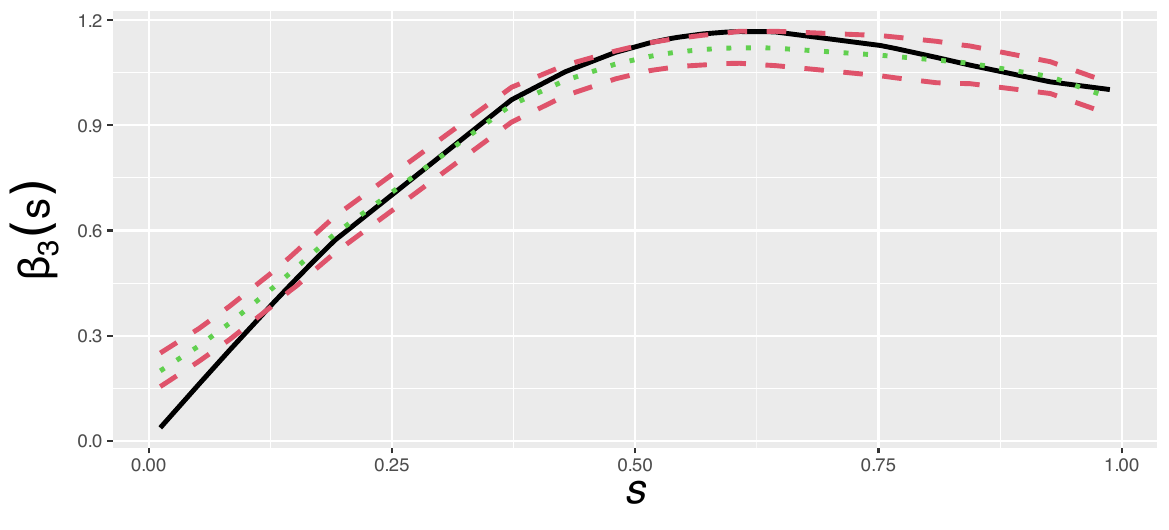}
	}
	\subfigure[]
	{
		\includegraphics[width=0.31\linewidth,height=0.31\linewidth]{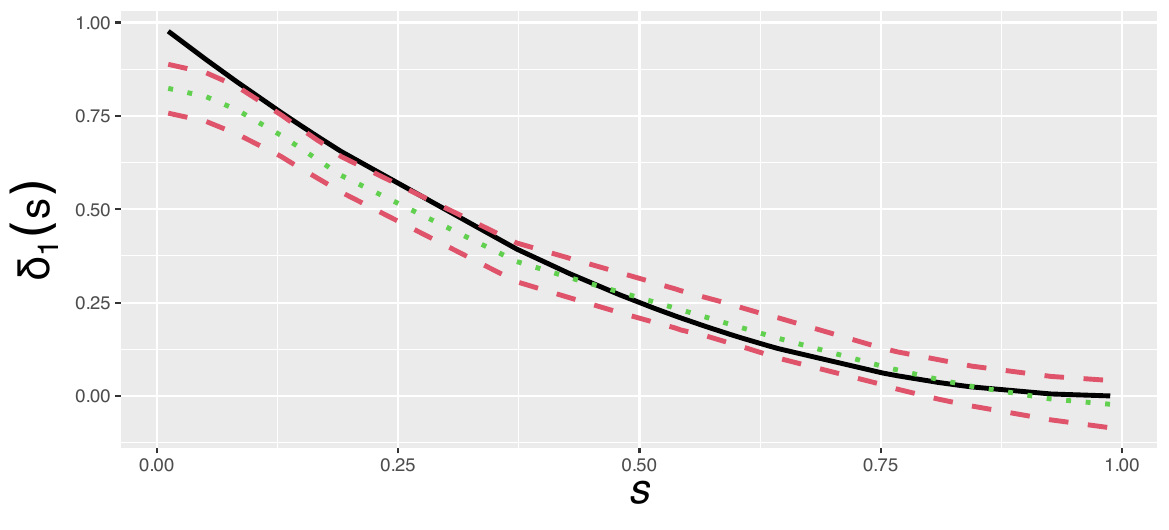}
	}
	\subfigure[]
	{
		\includegraphics[width=0.31\linewidth,height=0.31\linewidth]{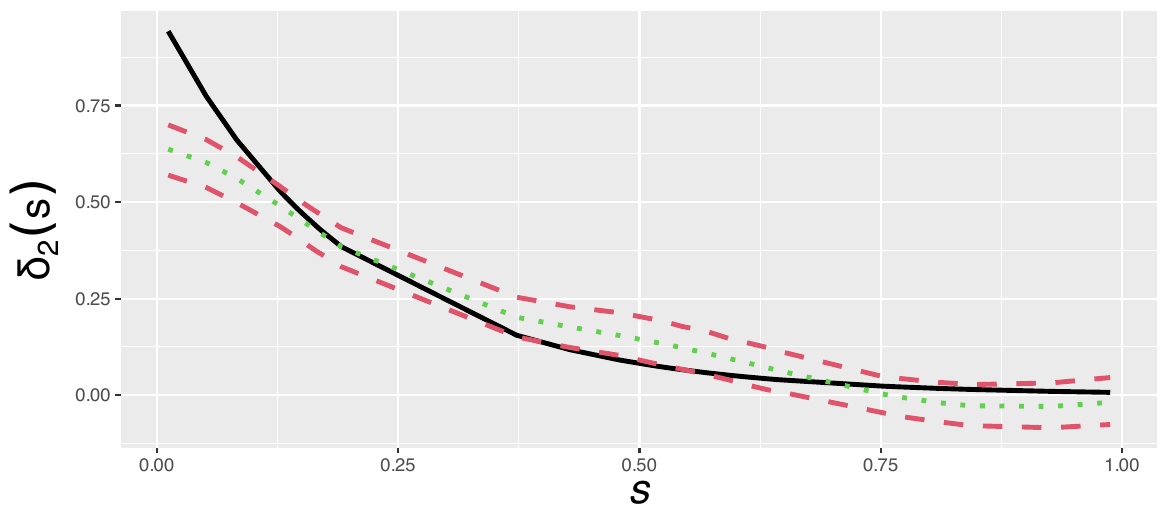}
	}
	\caption{(a)-(e) show the true function (solid line), the RKHS estimator (dashed line), and the 95\% pointwise confidence bands (long dashed line) of each component function.}
	\label{fig3}
\end{figure}

Set $n=100,200,400$ and $M=10, 30$, and take $1000$ repetitions. Table \ref{tab1} shows the bias and standard deviation of grouping parameter ${\bm \gamma}$.
It can be found from Table \ref{tab1} that the WLS estimators achieve smaller estimation errors compared with the LS estimators. 
Table \ref{tab2} reports the accuracy rate and Figure \ref{fig4} shows the boxplot results. Noting that the accuracy rates tend to one as $n$ increases, we conclude from Table \ref{tab2} and Figure \ref{fig4} that the WLS estimators perform better than LS estimators.
Table \ref{tab3} shows the RASEs and its standard deviation (in bold) of coefficient function estimators. It is clear that for all components of ${\bm \beta}(\cdot)$ and ${\bm \delta}(\cdot)$, 
the RASEs of the LS estimators and the WLS estimators decrease as $n$ increases, which verifies the theoretical consistency, and the WLS estimators achieve smaller estimation errors compared with the LS estimators.
Set $n=400$, $m=30$, Figure \ref{fig3} shows the true function (solid line), the RKHS estimator (dashed line), and the 95\% pointwise confidence bands (long dashed line) of each component functions.
It can be seen that for each component function, the estimated function is close to the true function, and the true function falls into their pointwise confidence bands except for the tail.

\begin{table}[!htbp]\tiny
	\renewcommand{\arraystretch}{1}
	\setlength{\tabcolsep}{2.6mm}
	\centering
	\setlength{\belowcaptionskip}{0.1cm}
	\caption{Bias and Standard Deviation (SD) of grouping parameter ${\bm \gamma}=(\gamma_1, \gamma_2)^\mathrm{T}$ in Section \ref{sec5.1}. LS stands for the initial estimation, and WLS stands for the weighted estimator.}
	\label{tab1}
	\resizebox{\textwidth}{!}{
		\begin{tabular}{cccccccccccccccc}
			\hline
			& & &\multicolumn{2}{c}{$n=100$}&  &\multicolumn{2}{c}{$n=200$}&  &\multicolumn{2}{c}{$n=400$}\\
			\cmidrule{4-5}
			\cmidrule{7-8} \cmidrule{10-11}\\ [-2mm]
			& & & $M=10$ & $M=30$   && $M=10$ & $M=30$  && $M=10$ & $M=30$  \\
			\hline
			\multirow{4}{*}{$\gamma_1$} & \multirow{2}{*}{LS} & Bias & 0.0982 & 0.0785 &&  0.0395 &0.0173  &&   0.0013 &  -0.0085 \\
			&  &SD & 0.5182 & 0.3672 &&0.2049  & 0.1407 && 0.0950 & 0.0833\\
			& \multirow{2}{*}{WLS}& Bias & 0.0763 &0.0697 && 0.0406 & 0.0273 && 0.0032  & 0.0001 \\
			& & SD & 0.4768 & 0.3451  && 0.2052 & 0.1391&& 0.0978 &0.0850 \\[6pt] 
			\multirow{4}{*}{$\gamma_2$} & \multirow{2}{*}{LS} & Bias & -0.0923 & -0.0565 && -0.0057  &0.0075  &&  0.0110  &  0.0255 \\
			&  &SD & 0.4546 &0.3081  && 0.1880 & 0.1129 &&0.0819 & 0.0715\\
			& \multirow{2}{*}{WLS}& Bias &-0.0755  & -0.0424&& -0.0056 & 0.0066 && 0.0121  & 0.0258 \\
			& & SD & 0.4185 &0.2924   && 0.1895 & 0.1115 &&0.0864  & 0.0734\\
			\hline
	\end{tabular}}
\end{table}

\begin{table}[!htbp]\tiny
	\renewcommand{\arraystretch}{1}
	\setlength{\tabcolsep}{2.6mm}
	\setlength{\belowcaptionskip}{0.1cm}
	\centering
	\caption{The accuracy rate of subgroup estimation in Section \ref{sec5.1}. LS stands for the initial estimation, and WLS stands for the weighted estimator.}
	\label{tab2}
	\resizebox{\textwidth}{!}{
		\begin{tabular}{cccccccccccccccc}
			\hline
			&& \multicolumn{2}{c}{$n=100$}&  &\multicolumn{2}{c}{$n=200$}&  &\multicolumn{2}{c}{$n=400$}\\
			\cmidrule{3-4}
			\cmidrule{6-7} \cmidrule{9-10}\\ [-2mm]
			& & $M=10$ & $M=30$   && $M=10$ & $M=30$  && $M=10$ & $M=30$  \\\hline
			Accuracy Rate & LS & 0.8856 & 0.9024 && 0.9369 & 0.9497 && 0.9688 & 0.9774 \\
			& WLS & 0.9802 & 0.9831 &&0.9875&0.9891 &&0.9918&0.9928\\
			\hline
	\end{tabular}}
\end{table}

\begin{table}[!htbp] \footnotesize
	\renewcommand{\arraystretch}{1}
	\setlength{\tabcolsep}{2.6mm}
	\centering
	\setlength{\belowcaptionskip}{0.1cm}
	\caption{The RASEs and its standard deviation (in bold) for each function in Section \ref{sec5.1}. LS stands for the initial estimation, and WLS stands for the weighted estimator.}
	\label{tab3}
	\resizebox{\textwidth}{!}{
		\begin{tabular}{cccccccccccccccc}
				\hline
				& & \multicolumn{2}{c}{$n=100$}&  &\multicolumn{2}{c}{$n=200$}&  &\multicolumn{2}{c}{$n=400$}\\
				\cmidrule{3-4}
				\cmidrule{6-7} \cmidrule{9-10}\\[-7pt]
				& & $M=10$ & $M=30$   && $M=10$ & $M=30$  && $M=10$ & $M=30$  \\
				\hline
				$\beta_1(\cdot)$&  LS & 0.0941 (\bfseries{0.0547}) & 0.0947 (\bfseries{0.0518}) &&  0.0679 (\bfseries{0.0377})& 0.0717 (\bfseries{0.0349}) &&  0.0531 (\bfseries{0.0246})  & 0.0560 (\bfseries{0.0224})  \\  
				&  WLS &  0.0719 (\bfseries{0.0405})& 0.0555 (\bfseries{0.0242}) && 0.0521 (\bfseries{0.0309})& 0.0451 (\bfseries{0.0179}) && 0.0406 (\bfseries{0.0229}) & 0.0392 (\bfseries{0.0137}) \\[6pt]
				$\beta_2(\cdot)$&  LS & 0.1525 (\bfseries{0.1012})&0.1350 (\bfseries{0.0910}) && 0.0973 (\bfseries{0.0641}) & 0.0909 (\bfseries{0.0585}) &&0.0643 (\bfseries{0.0381}) &  0.0641 (\bfseries{0.0389}) \\ 
				&  WLS & 0.0769 (\bfseries{0.0423})  &0.0532 (\bfseries{0.0256}) &&0.0550 (\bfseries{0.0285}) &0.0443 (\bfseries{0.0205})&& 0.0404 (\bfseries{0.0211}) &0.0368 (\bfseries{0.0157}) \\[6pt]
				$\beta_3(\cdot)$&  LS & 0.1686 (\bfseries{0.1021}) & 0.1609 (\bfseries{0.0879})&&0.1120 (\bfseries{0.0557})&0.1139 (\bfseries{0.0497})&& 0.0879 (\bfseries{0.0358}) & 0.0909 (\bfseries{0.0319}) \\
				&  WLS & 0.0917 (\bfseries{0.0461})& 0.0826 (\bfseries{0.0314}) &&0.0740 (\bfseries{0.0342})& 0.0749 (\bfseries{0.0243})&& 0.0628 (\bfseries{0.0288})& 0.0694 (\bfseries{0.0195}) \\[6pt]
				$\delta_1(\cdot)$&  LS &0.2285 (\bfseries{0.1624})  &0.2072 (\bfseries{0.1445}) && 0.1496 (\bfseries{0.0992})& 0.1324 (\bfseries{0.0852})&& 0.0932 (\bfseries{0.0545})&0.0932 (\bfseries{0.0504}) \\
				&  WLS &  0.0939 (\bfseries{0.0454}) & 0.0740 (\bfseries{0.0289}) && 0.0747 (\bfseries{0.0371}) &0.0616 (\bfseries{0.0224}) && 0.0567 (\bfseries{0.0272})&0.0542 (\bfseries{0.0184}) \\[6pt]
				$\delta_2(\cdot)$&  LS & 0.2390 (\bfseries{0.1487})&0.2319 (\bfseries{0.1352}) &&0.1615 (\bfseries{0.0873}) & 0.1618 ( \bfseries{0.0784})&&0.1178 (\bfseries{0.0514})&0.1201 (\bfseries{0.0447}) \\
				&  WLS & 0.1000 (\bfseries{0.0413}) &0.0938 (\bfseries{0.0263}) &&0.0813 (\bfseries{0.0326}) & 0.0861 (\bfseries{0.0210}) &&0.0693 (\bfseries{0.0268}) &0.0792 (\bfseries{0.0155}) \\
				\hline
		\end{tabular}}
	\end{table}

	\subsection{Subgroup Testing Model}\label{sec5.2}
	In this subsection, we evaluate the performance of the proposed test statistic. The data is generated from
	\begin{align*}
		Y_i(s_m) = {\bm X}_i^\mathrm{T}{\bm \beta}(s_m) + cn^{-1/2}{\bm \tau}(s_m)^\mathrm{T} \tilde{\bm X}_i I(Z_{1i} + {\bm Z}_{2i}^\mathrm{T} {\bm \gamma} > 0) + \nu_{i}(s_m) + e_i(s_m),
	\end{align*}
	where $c=(0,0.3,0.5,0.7,0.9,1.1,1.3)^\mathrm{T}$ with $c=0$ represents the null hypothesis, and ${\bm \gamma}=({\bm \gamma}_1, {\bm \gamma}_2)^T$.
	The predictors ${\bm X}_i $, $\tilde{\bm X}_i$, $Z_{1i}$, ${\bm Z}_{2i}$, the coefficient function ${\bm \beta}(\cdot)$,  ${\bm \delta}(\cdot)$, and the error structure are the same as Section \ref{sec5.1}.

	To compute the size and power of test statistic $T_{n,\max}$, we set $B=1000$ and $Q=1000$ in Algorithm \ref{alg3}. 
	Let $a$ be equally spaced numbers from 0.2 to 0.8.
	We adopt ${\bm \gamma}_1$ as the negative of the $a$-percentile of $Z_1+{\bm \gamma}_2{{\bm Z}_{2i}^*}$ such that ${\bm Z}_i^\mathrm{T}{\bm \gamma}$ divides the population into two groups with $a\%$ and $(1-a)\%$ observations.

\begin{figure}[!htbp]
		\centering
		\subfigure{\includegraphics[width=0.4\linewidth]{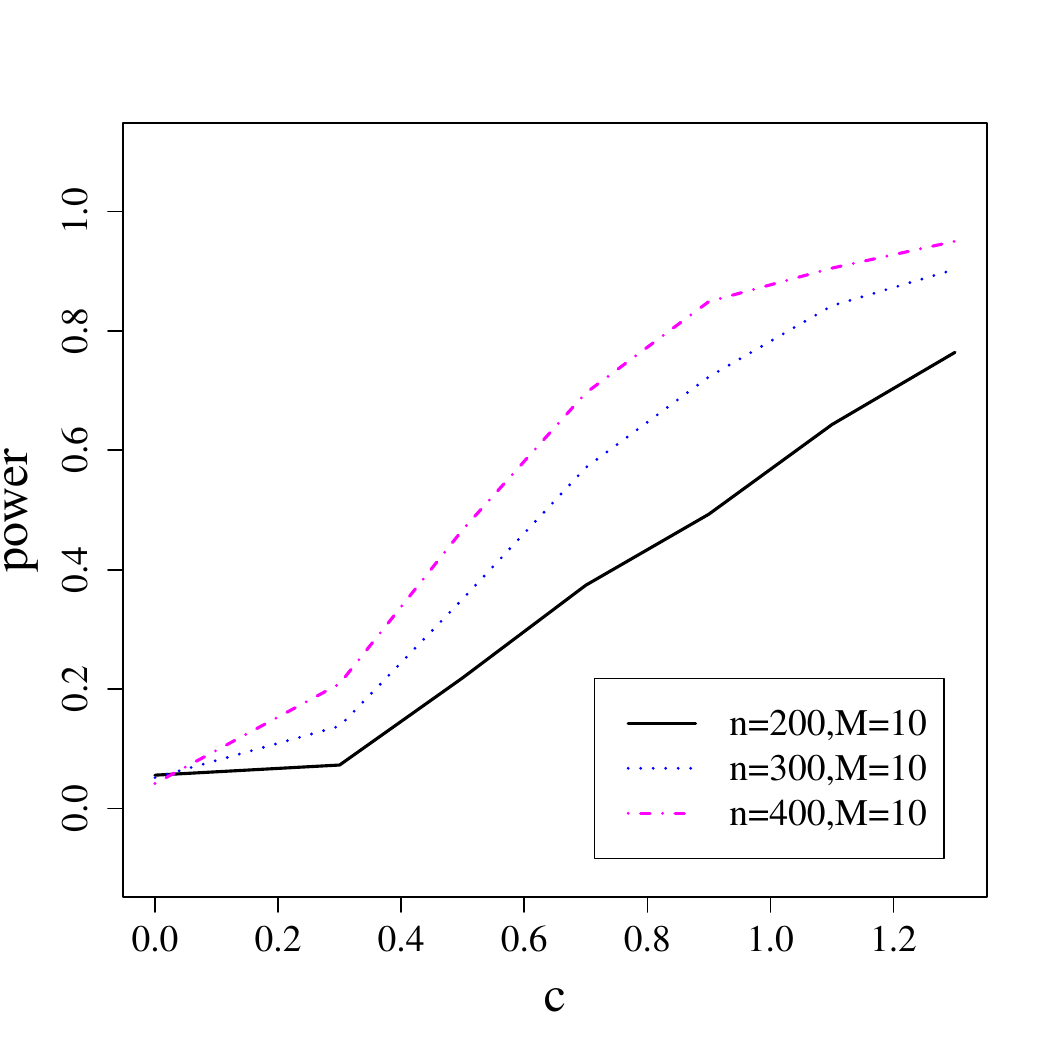}}\hspace{12pt}
		\subfigure{\includegraphics[width=0.4\linewidth]{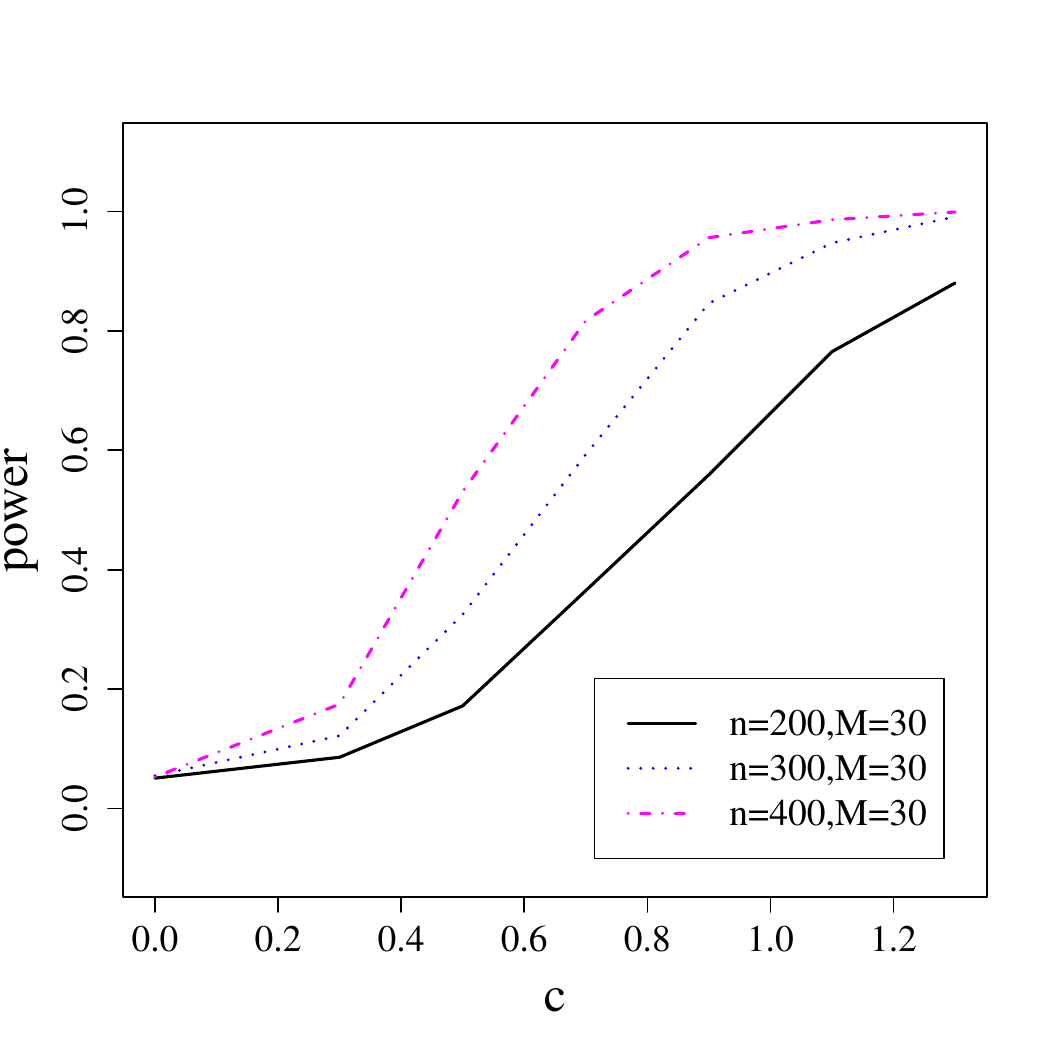}}
		\caption{The plot of the power of the test statistic $T_n$ in Section \ref{sec5.2} at nominal level 0.05 for different $c$ and different sample sizes $n,M$.}
		\label{fig5}
	\end{figure}

	Figure \ref{fig5} show the power of the test statistic at the nominal level $\alpha=0.05$ with varying $c$ and different sample size $n$ and a number of locations $M$, where 1000 replications are drawn. 
	From Figure \ref{fig5}, it can be found that when $c=0$, the estimated sizes are close to the corresponding nominal level under all scenarios.
	Moreover, the power increases to one when deviating from the null hypothesis. For visualisation purposes, Figure \ref{fig5} shows the power of the test statistic, and it is clear that the power increases to one with increasing $c$.

	\section{Applications}\label{sec6}
	In this section, we demonstrate the performance of the proposed method by analysing the mortality rate of the COVID-19 dataset and the air quality index (AQI) of Chinese cities. Due to limit of space, we put the AQI study in the Supplementary Material.
	
	Herein, we apply the change-plane analysis to the COVID-19 dataset mentioned in Section \ref{sec1} to further identify and test their subgroup structures.
	This dataset is obtained from the official website of the World Health Organization, which collected the mortality rate, the population aging and the medical care conditions in 137 countries for 120 days. 
	
	Our objective is to explore the relationship between the mortality rate and several scalar covariates, including the demographic features such as the Human development index ($x_1$); the number of people divided by land area (in square kilometers, $x_2$); the diabetes prevalence among people aged 20-79 (by percentage, $x_3$); the death rate from cardiovascular disease ($x_4$); the proportion of population aged 65 and above ($x_5$), and the socio-economic covariates such as the gross domestic product (GDP) per capita ($z_1$); the number of doctors per 1000 people ($z_2$); the number of hospital beds per 1000 people ($z_3$); the number of nurses per 1000 people ($z_4$).
	We consider the following functional change-plane model:
	\begin{align*}
		Y_i(s_m) = {\bm X}_i^\mathrm{T} {\bm \beta}(s_m) + \tilde{X}_i{\delta}(s_m) I(Z_{1i} + {\bm Z}_{2i}^\mathrm{T} {\bm \gamma} > 0) + \nu_{i}(s_m) + e_i(s_m),
	\end{align*}
	where $Y_i(s_m)$ is the COVID-19 mortality rate at the $m$th day since 100 confirmed cases for the $i$th country with $i=1,\cdots,137$ and $m=1,\cdots,120$, ${\bm X} = (1, x_1, x_2, x_3, x_4, x_5)^\mathrm{T}$, $\boldsymbol{\beta}(s) =(\beta_0(s),\beta_1(s),\cdots,\beta_5(s))^\mathrm{T}$, $\tilde{X} = x_5$, $Z_{1} = z_1$, ${\bm Z}_{2} = (1, z_2, z_3,z_4)^\mathrm{T}$. All variables are standardized. 
	
\begin{figure}[!htpb]
		\centering
			\includegraphics[height=10cm,width=\textwidth]{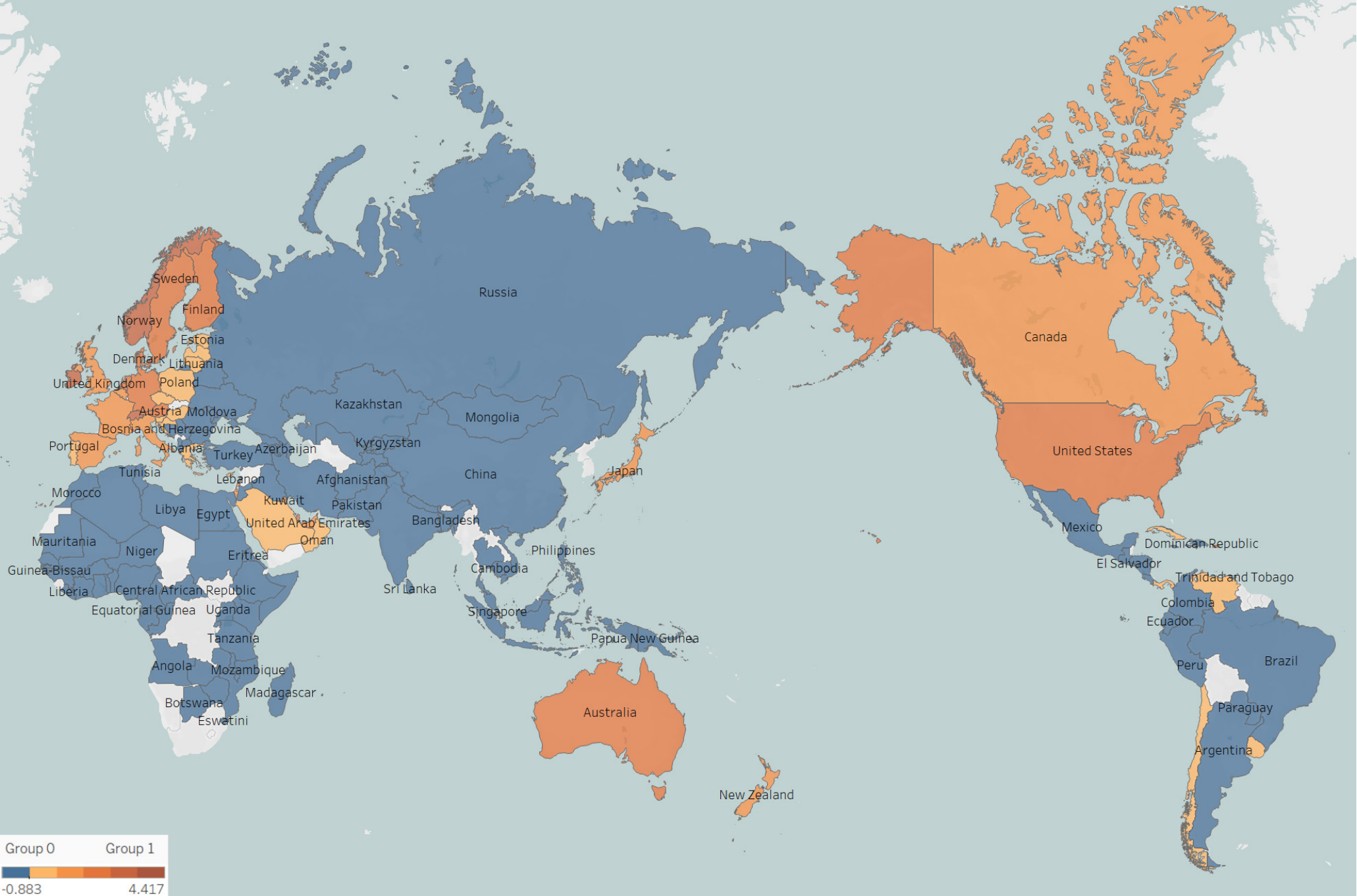}	
			\caption{Country grouping chart of the COVID-19 data: Group 0 (blue), Group 1 (orange), and missing samples (white). 
				The color is darker with larger values of $|Z_{1i}+\boldsymbol{Z}_{2i}^\mathrm{T}\hat{\boldsymbol{\gamma}}|$, indicating a higher probability to classify the country into the corresponding group.}
			\label{fig:covid19_country}
		\end{figure}

\begin{figure}[!htpb]
			\centering
			\includegraphics[width=0.6\linewidth]{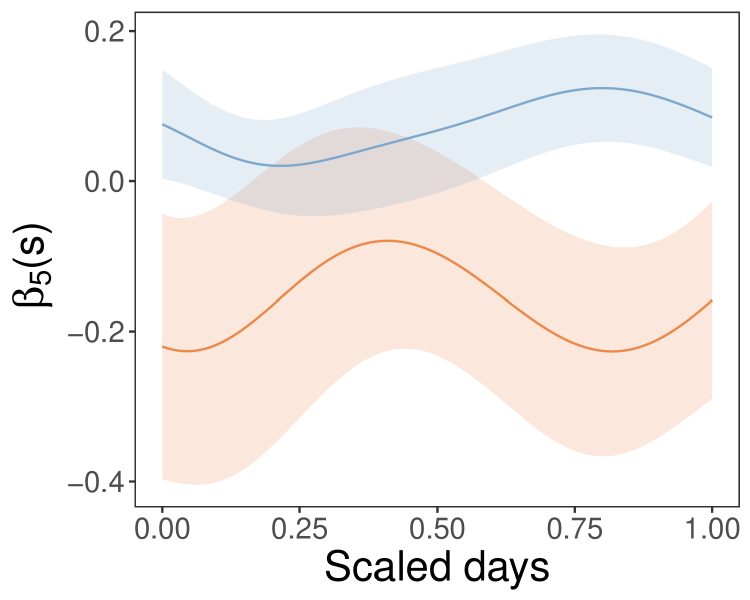}
			\caption{The estimated function $\hat{\beta}_5(\cdot)$ and its 95\% pointwise confidence intervals, where the orange line represents group 1 and the blue line represents group 0 in the COVID-19 data.}
			\label{fig:beta_hat}
		\end{figure}

	To test the existence of subgroups, we conduct the following hypothesis test
	$$H_0: {\delta}(s)=0, \text{ for all } s\in[0,1]\quad\text{ versus }\quad H_1: {\delta}(s)\neq0,\text{ for some } s\in[0,1].$$ 
	The resulting p-value is 0.002 by using the proposed test statistic, 
	which indicates that there is heterogeneity in countries caused by 
	the socio-economic covariates and the proportion of population aged 65 and above.
	By using the proposed estimation procedure, the estimated grouping parameters are $\hat{\bm \gamma}=(-0.0440, -0.0463, 0.0274, 0.1098)^\mathrm{T}$. Therefore, we can divide the 137 countries into two subgroups by the indicator function $I(Z_{1i}+\boldsymbol{Z}_{2i}^\mathrm{T}\hat{\boldsymbol{\gamma}}>0)$:  the baseline Group 0 ($I(\cdot)=0$) and the enhanced Group 1 ($I(\cdot)=1$), with 93 countries in Group 0 and 44 countries in Group 1. The mortality rates of two subgroups are shown in Figure \ref{fig1}. It is clear that the countries in Group 1 encountered higher mortality rates during the initial COVID-19 outbreak.
	Figure \ref{fig:covid19_country} shows the country groupings through a geographical map. 
	Figure \ref{fig:beta_hat} illustrates the estimated function $\hat{\beta}_5(\cdot)$ for two subgroups, highlighting that the influence of the elderly population proportion on mortality rates differs across subgroups.
	
\section{Discussion}\label{sec7}
This paper systematically investigates the application of change-plane analysis to functional data. Subgroups are well-defined and easy to interpret in this approach. We specifically consider a change-plane model for functional responses with scalar covariates. We introduced two algorithms for estimating functional parameters and grouping parameters in the RKHS framework, which took sufficiently into account the dependence of functional data. We rigorously established the asymptotic properties of the proposed estimators in the Sobolev space equipped with a proper inner product. To perform the inference, we proposed a novel supremum of squared score test statistic to assess the existence of subgroups and established the asymptotic properties of this test statistic. Extensive simulation studies provided empirical evidence that the estimation algorithm performs efficiently and the test statistic exhibits sound statistical power.

A promising direction for future research is to explore additional functional models, such as the function-on-function model or the functional linear model. Another idea for future work is to consider the multiple subgroups analysis. Our methodology could be easily extended to multiple subgroups in functional data, yet computational challenges may arise.




\bigskip
\begin{center}
	{\large\bf SUPPLEMENTARY MATERIAL}
\end{center}

The Supplementary Material includes additional simulation results, a case study and the proofs for the theorems in the article.


\bibliographystyle{agsm}
\bibliography{funcp-ref}

\end{document}